\documentclass[twocolumn,preprintnumbers,amsmath,amssymb,nofootinbib]{revtex4}

\usepackage{graphicx}
\usepackage{dcolumn}
\usepackage{bm}

\usepackage{epstopdf}
\def\bea{\begin{eqnarray}}
 \def\eea{\end{eqnarray}}

\def\pp{\mbox{$p$-$p$}}
\def\auau{\mbox{Au-Au}}

\def\pbpb{\mbox{Pb-Pb}}
\def\ppb{\mbox{$p$-Pb}}
\def\aa{\mbox{A-A}}
\def\pa{\mbox{$p$-A}}

\def\nn{\mbox{N-N}}

\def\pt{$p_t$}
\def\v2{$v_2$}
\def\yt{$y_t$}
\def\nch{$n_{ch}$}
\def\mt{$m_t$}

\def\mmpt{$\bar p_t$}

\begin{document} 

\preprint{Version 4.3}

\title{Quadrupole spectra derived from 2.76 TeV Pb-Pb identified-hadron $\bf v_2(p_t)$ data
}

\author{Thomas A.\ Trainor}\affiliation{University of Washington, Seattle, Washington 98195}


\date{\today}

\begin{abstract}

$p_t$-differential quantity $v_2(p_t)$ is meant to measure elliptic flow manifested by a dense QCD medium formed in high-energy nucleus-nucleus collisions. Elliptic flow may be referred to more neutrally as a cylindrical quadrupole component of the transverse motion of particle sources within a collision. As defined, $v_2(p_t)$ relies on an implicit assumption that almost all produced particles emerge from a single source. This article describes a detailed study of the algebraic structure of $v_2(p_t)$. A procedure is developed to derive a common monopole boost (``radial flow'') value and quadrupole $p_t$ spectra for several hadron species. The method is applied to $v_2(p_t)$ data for four hadron species from 2.76 TeV Pb-Pb collisions.  According to available $v_2(p_t)$ data the assumption of a single dominant particle source within A-A collisions is unjustified. Combined with a previous study of quadrupole amplitude variation for 200 GeV $p$-$p$ collisions these results demonstrate that quadrupole structure is related to a novel QCD process separate from projectile-nucleon dissociation and jet production. Given quadrupole evolution it is unlikely that a hydrodynamic description is relevant to that process.

\end{abstract}

\maketitle

 \section{Introduction} \label{intro}

In data from high-energy nucleus-nucleus (\aa) collisions an {\em azimuth quadrupole} component of 2D angular correlations on $(\eta,\phi)$ denoted by symbol $v_2$ has been attributed to elliptic flow. That assignment can be related to a statement in Ref.~\cite{olli}:\ ``We show that anisotropies in transverse-momentum distributions provide an unambiguous signature of transverse collective flow in ultrarelativistic nucleus-nucleus collisions.'' As such, $v_2$ data have played a central role in supporting arguments claiming achievement of a quark-gluon plasma (QGP)~\cite{perfect,qgp1,qgp2,keystone}. According to a conventional flow narrative elliptic flow should be sensitive to the early stage of high-energy \aa\ collisions where quarks and gluons are believed to be more-likely degrees of freedom. Correlation data demonstrating the presence of elliptic flow might thus confirm large energy and matter densities and copious parton rescattering to achieve a thermalized QGP~\cite{hydro2}. 

The RHIC experimental program has seemed to provide strong evidence confirming what may be called a flow-QGP narrative based on data obtained with certain preferred measures and techniques. It was therefore concluded in 2005 that a ``strongly-coupled QGP'' or ``perfect fluid'' is formed in central \auau\ collisions at RHIC energies~\cite{perfect}. However, one should distinguish between (a) the conjectured mechanism of elliptic flow and (b) the observed phenomenon of a cylindrical quadrupole\footnote{as in multipole expansion of sound emitted from a long cylinder.} on azimuth near midrapidity. The existence of (a) might imply (b) but observation of (b) does not {\em require} (a) and other observations may falsify (a). The present study applies novel analysis methods to recent LHC $v_2(p_t,n_{ch})$ data for identified hadrons. Some analysis results appear to contradict essential elements of a flow-QGP narrative.

Extensive studies of two-dimensional (2D) angular correlations~\cite{axialci,anomalous,multipoles,ppquad} have established that there are two main contributions to an observed azimuth quadrupole: (a) a nonjet (NJ) quadrupole component and (b) a jet-related quadrupole contribution derived from a same-side (on azimuth $\phi$) 2D jet peak representing {\em intra}\,jet angular correlations. Contribution (b) is often referred to as ``nonflow'' without acknowledging the dominant jet mechanism. 

{ The NJ quadrupole may be isolated accurately from jet-related and other contributions by model fits to 2D angular correlations~\cite{anomalous,davidhq}.  However, there has been a number of $v_2$ ``methods'' denoted $v_2\{\text{method}\}$ applied to data analysis, including $v_2\{2\}$ (two-particle), $v_2\{4\}$ (four-particle), $v_2\{\text{EP}\}$ (event-plane), $v_2\{\text{LYZ}\}$ (Lee-Yang zeros) and $v_2\{\text{SP}\}$ (scalar product). Methods are described as having varying sensitivities to ``nonflow'' biases.

Although quadrupole or $v_2$ data have played a central role in claims of QGP formation, hadron production near midrapidity appears to be dominated by two other mechanisms according to data: (a) longitudinal projectile-nucleon dissociation (soft) and (b) transverse large-angle parton scattering with jet formation (hard). The two mechanisms form the basis for a two-component (soft + hard) model (TCM) of hadron yields, spectra and correlations~\cite{ppquad,hardspec,anomalous,aliceptfluct,alicetomspec}. The TCM then provides an essential context for interpretation of NJ quadrupole data.

To better understand the relation between the NJ quadrupole and hydrodynamic (hydro) theory expectations for flows (Sec.~\ref{quadspecmeth}), $v_2(p_t,n_{ch})$ data for three species of identified hadrons from 200 GeV \auau\ collisions were processed to obtain their {\em quadrupole spectra}~\cite{quadspec} (Sec.~\ref{200gevquad}). The quadrupole spectra were found to be consistent with emission from a common boosted (i.e.\ moving)  source. The inferred narrow boost distribution {then} suggested emission from an expanding thin cylindrical shell~\cite{hydro2}. Three spectra were found to be equivalent modulo rescaling by factors consistent with a statistical model of hadron abundance~\cite{statmodel}. Quadrupole-spectrum parameters were quite different from those for single-particle (SP) spectra for most hadrons produced within the same collisions. The study concluded that the NJ quadrupole may be independent of most hadrons and represent a unique mechanism unrelated to a flowing  medium.

In the present study quadrupole spectra are obtained for 2.76 TeV \pbpb\ collisions (Sec.~\ref{quadspec}). Significant variation of monopole boost $\Delta y_{t0}$ with \nch\ event\ class is observed, but collision-energy variation of $v_2(p_t)$ is negligible. Major results are as follows: (a) Variation of the quadrupole {\em correlation amplitude} (correlated pair number) with event \nch\ over six orders of magnitude is observed for \pp\ {\em and} \aa\ collisions {\em combined}, revealing underlying simplicity (Sec.~\ref{multcent}). (b) The quadrupole amplitude increases with collision energy by two orders of magnitude from SPS to LHC energies whereas conventional measure $v_2$ appears to saturate at a constant value above 50 GeV (Sec.~\ref{quadedep}). The overall results suggest that the azimuth quadrupole source corresponds to novel QCD {\em color quadrupole} radiation that may be compared with dijet production as color dipole radiation. It is unlikely that a hydrodynamic description is relevant.

This article is arranged as follows:
Section~\ref{quadspecmeth}  defines  methods specific to inference of azimuth quadrupole spectra.
Section~\ref{200gevquad} reviews a previous quadrupole spectrum analysis of 200 GeV \auau\ data.
Section~\ref{quadspec} extends quadrupole spectrum analysis to data from 2.76 TeV \pbpb\ collisions.
Section~\ref{edep} reviews quadrupole-spectrum multiplicity and collision-energy trends.
Section~\ref{syserr} discusses systematic uncertainties.
Sections~\ref{disc1} and~\ref{summ}  present discussion and summary.
Appendix~\ref{boost} reviews the relativistic kinematics of boosted hadron sources and
Appendix~\ref{spspec} describes single-particle identified-hadron spectra used in the present study.

\section{Quadrupole spectrum methods} \label{quadspecmeth}

Reference~\cite{poskvol} provides an early analysis framework for flow measurement at the RHIC and LHC derived from the context of nucleon flow as observed at the Bevalac~\cite{bevalac}. At lower energies most of the collision system may participate in collective motion of nucleons and there is no significant jet production. In that case one might assume emission from a flowing nucleon distribution as the only particle source. Almost all emitted nucleons may then share a common boost distribution. It  follows from Eq.~(1) of Ref.~\cite{poskvol}  that
\bea \label{v2av} 
v_2(p_t) &=&  \frac{\int_0^{2\pi} d\phi_r  \cos(2\phi_r) \bar \rho(p_t,\phi_r)}{\int_0^{2\pi} d\phi_r  \bar \rho(p_t,\phi_r)},
\\ \nonumber
&\rightarrow & \langle  \cos(2\phi_r) \rangle(p_t)
\eea
where $\bar \rho(p_t,\phi_r)$ represents a total particle density averaged over a small $\eta$ interval  at midrapidity and $\phi_r \equiv \phi - \Psi_r$ is relative to some reaction-plane angle $\Psi_r$. {  The assumption of an {\em event-wise} reaction- or event-plane angle $\Psi_r$ relating to almost all produced particles and to be {\em estimated from particle data} is fundamental to that approach. Reference~\cite{njquad} demonstrates in its Sec.~IV B that the event-plane \{EP\} method reduces to a manipulation of two-particle correlations that does not require an actual reaction plane or an event-plane estimate.}
{Note that a number of methods emphasizing cumulant analysis~\cite{cumulant} and/or subevents~\cite{subevents} does not require explicit estimation of a reaction plane.}
{However,  $v_2$ measurement scenarios related to Eq.~(\ref{v2av}) assume a single dominant ``quadrupole'' hadron production mechanism represented there by $\bar \rho(p_t,\phi_r)$.} More-recent analysis of RHIC and LHC data reveal two dominant {\em nonquadrupole} production mechanisms (nucleon dissociation and dijets) that exhibit no significant evidence for source boosts. Conventional $v_2$ measures thus appear to conflate three hadron production mechanisms as discussed  below.

\subsection{$\bf v_2(p_t)$ at RHIC and LHC energies}

At RHIC and LHC energies participant nucleons dissociate into hadron fragments, and there is copious jet production arising from parton binary collisions within \nn\ collisions. That description runs counter to what might be described as a flow-QGP narrative in which almost all produced hadrons emerge from a dense flowing partonic medium (QGP) extrapolated from the Bevalac scenario and represented by monolithic density $\bar \rho(p_t,\phi_r)$. The density relevant to the denominator of Eq.~(\ref{v2av}) is different from the numerator, and the second line of Eq.~(\ref{v2av}) is  therefore not valid. An alternative approach is required: numerator and denominator are considered separately.

What one observes for the denominator of Eq.~(\ref{v2av}), the azimuth average of total particle density $\bar \rho(p_t,\phi_r)$, may be described as a {\em monopole} spectrum $\bar \rho_0(p_t)$ represented by a two-component (soft+hard) model (TCM) that has been applied  to many collision systems (e.g.\ \cite{ppprd,ppbpid,tompbpb}),
\bea \label{rho0}
\bar \rho_0(p_t) &=&(N_{part}/2)\bar \rho_{sNN}(p_t) + N_{bin} \bar \rho_{hNN}(p_t).~~~
\eea
$\bar \rho_{sNN}$ and $\bar \rho_{hNN}$ are soft and hard (jet) particle densities for \nn\ ($\approx$ \pp) collisions.
$N_{part}/2$ (participant-nucleon pairs) and $N_{bin}$ (\nn\ binary collisions) describe \aa\ collision geometry.  Monopole spectra exhibit no significant transverse boost (radial flow) for any system~\cite{noblast}.

The numerator of Eq.~(\ref{v2av}), $V_2(p_t) \equiv \bar \rho_0(p_t) v_2(p_t)$, is a Fourier amplitude corresponding to a $\cos(2\phi)$ term and derived from 2D angular correlations on $(\eta,\phi)$. At RHIC and LHC energies jets contribute strongly to such correlations. Different correlation analysis methods (e.g.\ as described in Ref.~\cite{poskvol}) may or may not distinguish between jet contributions (called ``nonflow'') and ``flow.'' 

For this article, angular correlations on $(\eta,\phi)$ are separated into a nonjet (NJ) quadrupole ($\cos(2\phi)$) component independent of $\eta$ and other structure dominated by jets. The NJ quadrupole component may be accurately isolated via model fits to 2D angular correlations~\cite{anomalous,ppprd}. An example is given in Fig.~\ref{fits} below. Those $v_2$ data are denoted by $v_2\{\text{2D}\}$ whereas other analysis methods are conventionally denoted by $v_2\{\text{method}\}$, e.g.\ $v_2\{2\}$ or $v_2\{4\}$. {   The relation $V_2\{\rm 2D\}(p_t) \equiv \bar \rho_0 v_2\{\rm 2D\}(p_t)$ defines the quantity in the numerator below.} Given  that context Eq.~(\ref{v2av}) may be reexpressed as
\bea \label{v2struct}
v_2(p_t) \hspace{-.05in} &=& \hspace{-.05in} \frac{V_2\{\rm 2D\}(p_t) + \text{jet contribution}}{(N_{part}/2)\bar \rho_{sNN}(p_t) + N_{bin} \bar \rho_{hNN}(p_t)}.~~~~
\eea
Eq.~(\ref{v2struct}) is more applicable above $\sqrt{s_{NN}} = 50$ GeV where the TCM accurately describes hadron and jet production within \nn\ collisions dominated by low-$x$ gluons.

For simplicity it is assumed that the collision system includes a simple boost field as expressed on transverse rapidity \yt\footnote{$y_{ti} \equiv \ln[(p_t + m_{ti})/m_i]$ for hadron species $i$ with mass $m_i$.} with monopole $\Delta y_{t0}$ and quadrupole $\Delta y_{t2}(\phi_r)$ boost components. In the scenario implicit for Eq.~\ref{v2av} one should expect a common flow field to be manifested in both numerator and denominator of Eq.~(\ref{v2av}). That is, if a monopole boost plays a significant role in the numerator then it should appear somehow in the azimuth-averaged denominator. But as noted above, differential analyses of identified-hadron \mbox{A-B} spectra $\bar \rho_0$ for RHIC~\cite{hardspec} and LHC~\cite{ppbpid,tompbpb,noblast} indicate no radial flow $\Delta y_{t0}$ component but do establish a substantial jet-related (hard)  component for all A-B systems. 
Given observations the spectrum integrated in the numerator is designated $\bar \rho_2(p_t,\phi_r,\Delta y_{t0},\Delta y_{t2})$ (quadrupole) and the azimuth-averaged denominator of Eq.~(\ref{v2av}) has been denoted $\bar \rho_0(p_t)$ (monopole). This analysis then probes the relationship between $\bar \rho_0$ and $\bar \rho_2$.
Inference of quadrupole spectra $\bar \rho_2$ from $v_2(p_t,n_{ch})$ data requires compatible SP spectra $\bar \rho_0(p_t,n_{ch}) \equiv d^2 n_{ch} / p_t dp_t d\eta$ and factorization of $V_2\{\rm 2D\}(p_t)$ via the Cooper-Frye formalism~\cite{cooperfrye,quadspec}. 

\subsection{Quadrupole spectrum definition}

A spectrum description for the NJ quadrupole component may be derived from experimental $v_2(p_t)$ data assuming that  
(a) the quadrupole component arises from one or more hadron sources each with an eventwise azimuth-dependent radial boost distribution $\Delta y_t(\phi_r)$, 
(b) a quadrupole spectrum may be nearly thermal {\em in its boost frame} 
and 
(c) quadrupole sources {\em may} produce only a fraction of the hadrons in a collision, independent of SP-spectrum $\bar \rho_0$ soft and hard components.
That description is consistent with a conjecture in Sec.~\ref{implications} that for more-central \aa\ collisions multiple independent quadrupole sources (three-gluon interactions) may be active.
Given those possibilities a $\phi_r$-dependent spectrum at midrapidity, for those hadrons associated with the NJ quadrupole component, may be modeled {\em in the lab frame} by
\bea \label{fullquad}
\bar \rho_2(m_t,\phi_r) \hspace{-.04in} &\propto& \hspace{-.04in}  \exp\{ -\left(\gamma_t(\phi_r)[m_t - \beta_t(\phi_r) p_{t} ] - m\right)  /T_2\} \nonumber
\\ 
\bar \rho_2(y_t,\phi_r) \hspace{-.04in} &\propto& \hspace{-.04in} \exp\{ -m[\cosh(y_t - \Delta y_{t}(\phi_r)) - 1]/T_2  \},
\eea
{where the first line is compatible with the style of Ref.~\cite{cooperfrye} and the second line is based on transverse rapidity \yt.}
A Boltzmann exponential on $m_{ti} = m_i \cosh(y_{ti})$ for a locally-thermal source is assumed for simplicity. See App.~\ref{boosteq} for relevant definitions.

Based on relativistic kinematics reviewed in App.~\ref{boost} and the assumed boost model expressed by  Eq.~(\ref{quadboost}) the spectrum defined by Eq.~(\ref{fullquad}) may be factored as
\bea \label{factor}
\bar \rho_2(y_t,\phi_r) \hspace{-.0in} &=& A_{2} \exp\{ -m[\cosh(y_t - \Delta y_{t0}) - 1]/T_2  \}\times 
 \nonumber \\
&& \hspace{-.2in} \exp[m'_t \, \{\cosh[ \Delta y_{t2}\, \cos(2  \phi_r)] - 1\}/T_2] \times 
 \nonumber \\
&& \hspace{-.2in} \exp\{p'_t \, \sinh[\Delta y_{t2}\, \cos(2  \phi_r)] /T_2\} 
 \nonumber \\
&\equiv& \bar \rho_2(y_t,\Delta y_{t0})\times F_1(y_t,\phi_r, \Delta y_{t2})\times 
 \nonumber \\
&&F_2(y_t,\phi_r ,\Delta y_{t2}),
\eea  
where primes indicate {\em momenta in the boost frame}.
The last line defines azimuth-dependent factors $F_1(y_t,\phi_r)$ and $F_2(y_t,\phi_r)$ in terms of monopole and quadrupole components of  radial boost. The objective is {\em azimuth-averaged} quadrupole spectrum  $\bar \rho_2(y_t,\Delta y_{t0})$ emitted from a conjectured boosted hadron source as one factor of Fourier amplitude $V_2(p_t)$ inferred from $v_2(p_t)$ measurements. 

Given azimuth-dependent spectrum $\bar \rho_2(y_t,\phi_r)$ defined by Eq.~(\ref{factor}),  its quadrupole-related Fourier amplitude is
\bea
V_2(y_t)
&=&\frac{1}{2\pi} \int_{-\pi}^{\pi} d\phi_r \ \cos(2 \phi_r)  \bar \rho_2(y_t,\phi_r) .
\eea
The full integral over factors $F_1$ and $F_2$ in Eq.~(\ref{factor}) is
\bea \label{fint}
 \frac{1}{2\pi} \int_{-\pi}^{\pi} \hspace{-.14in} d\phi_r F_1(y_t,\phi_r) F_2(y_t,\phi_r)\cos(2  \phi_r)  
\hspace{-.05in } &=& \hspace{-.05in}p'_t \frac{\Delta y_{t2}}{2T_2}f(y_t),~~~~
\eea
where $f(y_t,\Delta y_{t0},\Delta y_{t2})$ is an $O(1)$ correction factor determined by ratio $\Delta y_{t2}/\Delta y_{t0}$: $f(y_t)$ remains closer to 1 the smaller is that ratio~\cite{quadspec}. Combining factors gives for each hadron species $i$
\bea \label{combfac}
V_{2}(y_{t},\Delta y_{t0},\Delta y_{t2}) \hspace{-.04in} &=&\bar \rho_{0}(y_t) v_{2}(y_t)  ~~{\rm data}
\\ \nonumber  
&\approx& 
 \hspace{-.04in}p'_t\, \frac{ \Delta y_{t2}}{2T_2} \, \bar \rho_{2}(y_t,\Delta y_{t0}).
\eea
establishing a direct relation between $v_{2i}(y_t)$ data and quadrupole spectra $\bar  \rho_{2i}(y_t,\Delta y_{t0})$. $p'_t$ is \pt\ in the boost frame, $T_2$ is the quadrupole-spectrum slope parameter, $\Delta y_{t0}$ is the monopole source boost, and $\Delta y_{t2}$ is the amplitude of the source-boost quadrupole modulation. {   $v_{2i}(y_t)$ data might conflict with that simple model to reveal a monopole source-boost {\em distribution} on $\Delta y_{t0}$ corresponding to Hubble-like expansion of a dense bulk medium.} {  Note that while $\phi_r = \phi - \Psi_r$ appears in  formulas for this subsection there is no apparent need to determine $\Psi_r$ experimentally, and a global reaction plane may be irrelevant. These formulas should be understood to apply to {\em individual quadrupole emissions} as discussed in Sec.~\ref{implications}}

\subsection{Inferring $\bf \bar \rho_2(y_t,\Delta y_{t0})$ from measured $\bf v_2(y_t)$ data}

A quadrupole spectrum may be inferred from measured quantities by the relation (for fixed monopole boost $\Delta y_{t0}$)
\bea \label{stuff}
\bar \rho_0(y_t)\, \frac{v_2(y_t)}{p_t} \hspace{-.05in} &=& \hspace{-.05in} \left\{\frac{p'_t}{p_t\, \gamma_t(1 - \beta_t)}\right\}\,  \left\{\frac{\gamma_t(1 - \beta_t)}{2T_2}\right\} \times \\ \nonumber
&&f(y_t,\Delta y_{t0},\Delta y_{t2})\,\Delta y_{t2}\,  \bar \rho_2(y_t,\Delta y_{t0}).
\eea
The quantities on the left are measured experimentally. Divisor $p_t$ is introduced on the left as a trial value and is later replaced by boost-frame $p_t'$ if boost $\Delta y_{t0}$ is inferred from data. $\bar \rho_2(y_t,\Delta y_{t0})$ on the right is the sought-after quadrupole spectrum. Power-law exponent $n_2$ and $T_2$ common to all hadron species may be estimated from the $\bar \rho_2(y_t,\Delta y_{t0})$ spectrum shape inferred from data as illustrated below. The first factor on the right, shown in Fig.~\ref{boost3} (right), is determined only by $\Delta y_{t0}$ and deviates from unity only near a zero intercept on \yt. The numerator of the second factor ($\approx 0.55$) is also determined by $\Delta y_{t0}$. Thus,  factors in the first line on the right and the {\em shape} of $\bar \rho_2(y_t,\Delta y_{t0})$ are determined by data on the left.

\section{200 $\bf GeV$ $\bf Au$-$\bf Au$ quadrupole spectra} \label{200gevquad}

Given the above framework, a corresponding procedure is applied to infer quadrupole spectra $\bar \rho_{2i}(y_{ti},n_{ch})$ from 200 GeV \auau\ $v_2(p_t)$ data for three hadron species. The procedure consists of a sequence of transformations that may be described as {\em homeomorphisms} which better present the information carried by $v_2$ data. Below is a  summary of a \auau\ analysis reported in Ref.~\cite{quadspec}. Note that certain data have been corrected based on recent information as summarized in the beginning of Sec.~\ref{modelfits}.

\subsection{NJ quadrupole $\bf v_2(p_t)$ data in two formats}

Fig.~\ref{x1} (left) shows 200 GeV $v_2(p_t)$ data for three hadron species vs $p_t$ in a conventional plotting format averaged over 0-80\% \auau\ centrality~\cite{v2pions,v2strange}.  The curves extending off the top edge of the panel are $v_2 \propto p_t'$ as in Eq.~(\ref{v2simple}) and Fig.~\ref{schema1} (left) reflecting expected ideal-hydro trends for a single boost value $\Delta y_{t0} = 0.6$ that {\em describe $v_2$ data} for $p_t < 2$  GeV/c. For Hubble-like expansion of a bulk medium the source boost distribution should be broad. The solid, dashed and dash-dotted curves passing through data are described in Sec.~\ref{describe}.
The solid triangles ($\pi$ new) represent more-recent  minimum-bias (MB) 200 GeV pion data with higher statistics~\cite{newstarpion} that were encountered after Ref.~\cite{quadspec} was published.

\begin{figure}[h]
    \includegraphics[width=1.63in]{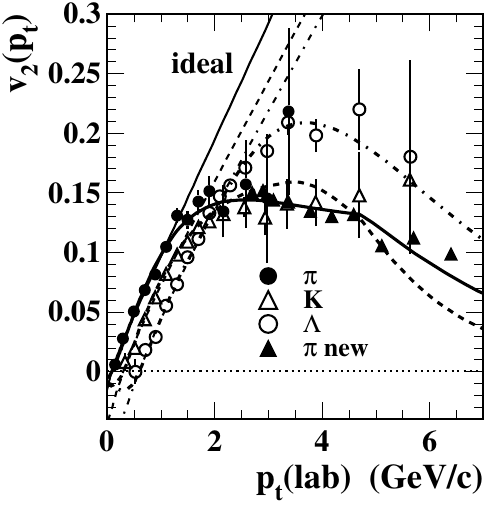}
     \includegraphics[width=1.67in]{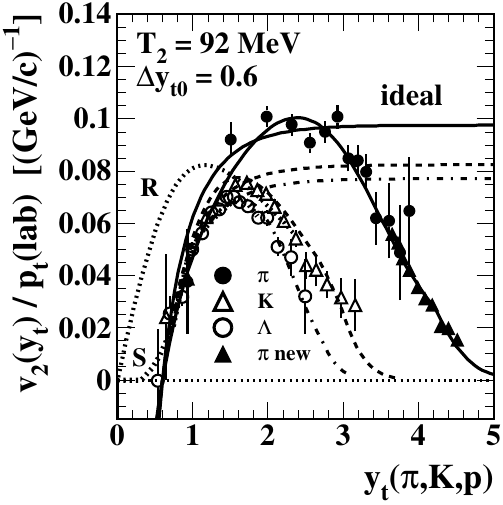}
\caption{ \label{x1}  
Left: $v_2(p_t)$ data for three hadron species plotted in a conventional format. The kaon and Lambda data are from 0-80\% central 200 GeV \auau\ collisions~\cite{v2strange}. Data representing pions (solid dots) are hadron data for 16-24\% 130 GeV \auau\ collisions scaled up by 1.2~\cite{v2pions}. Newer 200 GeV pion data extending beyond 6 GeV/c (solid triangles) are from Ref.~\cite{newstarpion}.
Right: The same data plotted as ratio $v_2(p_t) / p_t(\text{lab})$ on transverse rapidity. The curve labeled R is an earlier viscous-hydro prediction for identified protons~\cite{rom}. {   The curve labeled S is a more-recent prediction for protons~\cite{shen}. See Fig.~\ref{x2} (left).} Solid, dashed and dash-dotted curves through data are described  in Sec.~\ref{describe}.
} 
\end{figure}

Fig.~\ref{x1} (right) shows the same data divided by $p_t(\text{lab})$ and plotted vs transverse rapidity $y_{ti}$ with proper mass $m_i$ for each hadron species $i$. It is notable that the data for three hadron species pass through a common zero intercept at $y_t \approx 0.6$ ($\rightarrow \Delta y_{t0}$) consistent with emission from an expanding thin cylindrical shell~\cite{hydro2}. The ``ideal'' curves approaching a constant value at larger $y_t$ correspond to the ``ideal-hydro'' trends from the left panel. {  The $v_2(p_t)$ data in contrast fall sharply away from the hydro trends. That falloff has led to inference of low $\eta/s$ (viscosity/entropy) values~\cite{luzrat} with claims of ``perfect fluid'' formation~\cite{perfect}. An explanation in terms of the  ratio structure of $v_2(p_t)$ in Eq.~(\ref{v2struct}) is presented in Sec.~\ref{hydropredict}.}

Fig.~\ref{x2} (left) shows an expanded view of Fig.~\ref{x1} (right) for Lambda baryons {  compared to an earlier viscous-hydro theory curve for protons (dotted curve R)~\cite{rom} and a more-recent curve (bold dotted S) from Ref.~\cite{shen} also for protons. The quadrupole source-boost distribution is best determined} in  this case by protons or Lambdas for two reasons: (a) For a given detector \pt-acceptance lower bound, data distributions on \yt\ extend to a lower value for more-massive hadrons since $y_{ti} \approx p_t / m_i$ at lower \pt. The vertical dotted line marks a lower limit for protons or Lambdas whereas the corresponding limit for pions is near $y_t = 1$. (b) Given $\Delta y_{t0} \approx 0.6$,  data from heavier hadrons with more-limited statistics would provide little additional information.
The open squares are recent Lambda data for 0-10\% \auau\ collisions~\cite{newstarpion} that follow a trend with significant negative values below the  intercept near $y_t = 0.6$ and confirm the dash-dotted trend {\em predicted} by Ref.~\cite{quadspec}. 
{  Viscous-hydro results S from the more-recent theory~\cite{shen} are discussed in Sec.~\ref{shen}.}

\begin{figure}[h] 
    \includegraphics[width=1.65in]{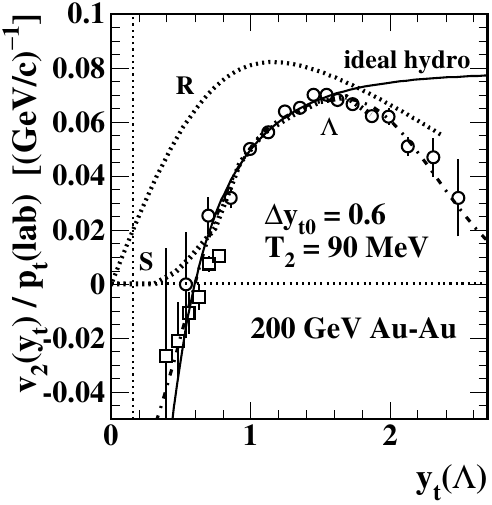}
   \includegraphics[width=1.64in]{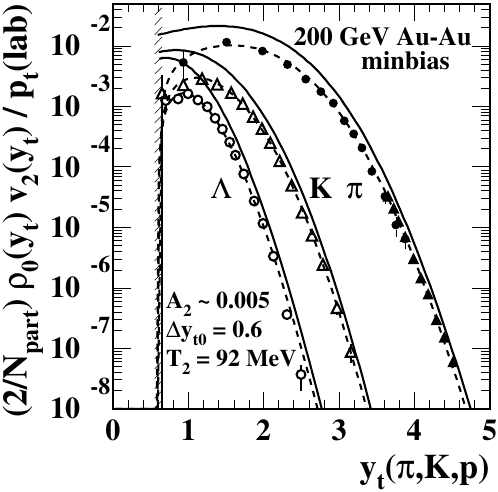}
\caption{\label{x2}  
Left: Lambda data from Fig.~\ref{x1} (right, open points) on an expanded \yt\ scale compared to older viscous-hydro theory curve R for protons~\cite{rom} {  and more-recent hydro result S (10-20\% central \pbpb)~\cite{shen} also for protons.}
The open boxes are more-recent Lambda data for 0-10\% central \auau\ collisions~\cite{newstarpion}. The dotted line marks a detector acceptance limit for charged hadrons at $p_t \approx 0.15$ GeV/c. The curve marked ``ideal hydro'' is simply proportional to $p_t'=p_t$(boost).
Right: Data from Fig.~\ref{x1} (right) multiplied by SP spectra in the form $(2/N_{part}) \bar \rho_0(y_t)$ derived from hadron spectra in Ref.~\cite{hardspec}. The curves are described in the text.
} 
\end{figure}

\subsection{Quadrupole spectra inferred from $\bf v_2(p_t)$ data} \label{describe}

Fig.~\ref{x2} (right) shows data (points) from Fig.~\ref{x1} (right) multiplied by SP spectra in the form $(2/N_{part})\bar \rho_{0i}(y_t)$ for each hadron species to obtain $(2/N_{part})V_{2i}(y_{ti}) / p_t(\text{lab})$. Solid triangles represent more-recent MB 200 GeV pion data with higher statistics~\cite{newstarpion} as in Fig.~\ref{x1}. That figure corresponds to Fig.~9 of Ref.~\cite{quadspec}.

The SP spectrum for 200 GeV pions is shown in Fig.~\ref{alice5} (a) of App.~\ref{spspec} for comparison with 2.76 TeV \pbpb\ spectra. Spectra for the lower energy are somewhat softer as expected. The data in Fig.~\ref{x2} (right) then include by construction a factor $p_t' / p_t$ relative to quadrupole spectrum $ \bar \rho_2(y_t,\Delta y_{t0})$ as defined by Eq.~(\ref{stuff}). Curves through data are back-transformed from a universal quadrupole spectrum on $m_{ti}' - m_i$ (solid curve in Fig.~\ref{xbig}) with (dashed) and without (solid) factor $p_t' / p_t$ derived from Eq.~(\ref{kine}) that includes factor $\gamma_t (1-\beta_t) \approx 0.55$ for $\Delta y_{t0} \approx 0.6$. 

Fig.~\ref{xbig} shows quadrupole spectra on  $m_{ti}' - m_i$ {\em in the boost frame} for three hadron species as defined by the $y$-axis label. The lab-frame quadrupole spectra (points) in Fig.~\ref{x2} (right) are multiplied by $p_t / p_t' = p_t(\text{lab}) / p_t(\text{boost})$ (since $\Delta y_{t0}$ is approximated from the data common zero intercept in Fig.~\ref{x1}, right), transformed to $y_t'$ in the boost frame by shifting data and curves to the left on \yt\ by $\Delta y_{t0}$  (hence $\Delta y_{t0} \rightarrow 0$), and transformed to densities on $m_{ti}' - m_i$ with Jacobian $y_{ti}' / (m_{ti}' - m_i) p_t'$. The $v_2/p_t$ data errors have been similarly transformed assuming that SP spectrum errors are negligible. The resulting spectra, rescaled relative to pions with statistical-model factors indicated on the plot, are found to coincide precisely over the entire $m_{ti}'$ acceptance. Note that Lambda data extend to $p_t = 5.6$ GeV/c in the lab frame but only 3 GeV/c in the boost frame (or $m_{ti}' - m_i = 2.25$ GeV/$c^2$). The new pion data (solid triangles) extend to 6.5 GeV/c in Fig.~\ref{x1} (left) and fall exactly on curves back transformed  from  $\hat S_2(m_{ti}')$ in Fig.~\ref{xbig} that was determined before the new data were encountered. The data cover a \pt\ interval far beyond what is conventionally assumed valid for hydro treatments (e.g.\ $p_t < 3$ GeV/c for Ref.~\cite{shen}).

\begin{figure}[h] 
   \includegraphics[width=3.3in]{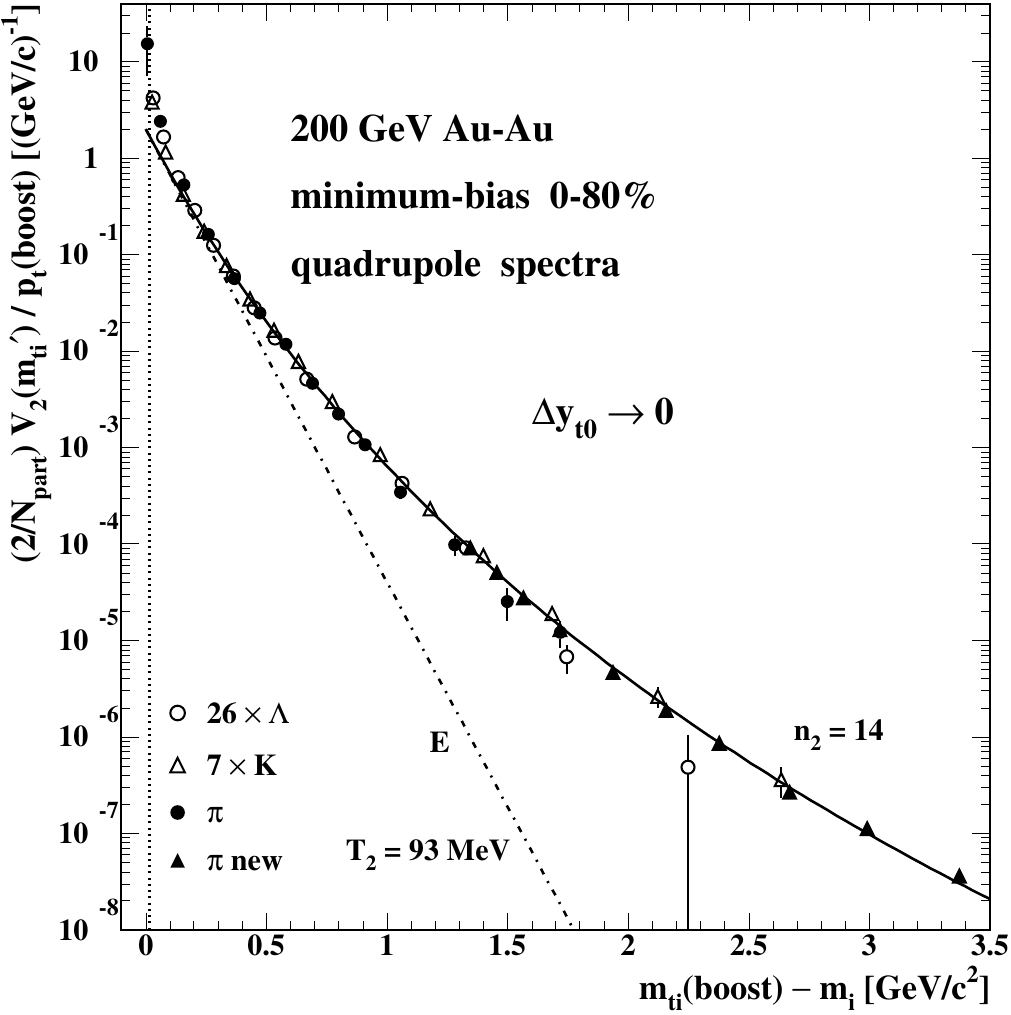}
\put(-85,83) {$\bf \hat S_2(m_t)$}
\caption{\label{xbig}  
Data from Fig.~\ref{x2} (right) divided by  kinematic factor $p_t' / p_t$ defined in Eq.~(\ref{kine}) and transformed to $m_t(\text{boost}) - m_i$. Those data, rescaled by the factors indicated on the plot, then coincide on a single locus modeled by distribution $\hat S_2(m'_t)$ (solid curve). $\hat S_2(m_t)$ spectrum model parameters are quite different from those for SP hadron spectra~\cite{ppprd,hardspec}. 
}  
\end{figure}

Solid curve $\hat S_2(m_{ti}')$ is an exponential with power-law tail having parameters $(n_2,T_2)$ indicated on the plot. As noted, that function is back-transformed to generate solid, dashed and dash-dotted curves through data in previous figures. Up to an overall constant, three numbers -- $\Delta y_{t0} \approx 0.6$, $T_2 \approx 93$ MeV and $n_2 \approx 14$ -- accurately describe all MB $v_2(p_t)$ data for three hadron species. Those hadrons associated with the NJ  quadrupole {   appear to follow a unique spectrum} representing not Hubble-like expansion of a bulk medium but {  rather a thin shell expanding with fixed speed}. The quadrupole spectrum is quite different from the SP \pt\ spectrum describing most hadrons. These data include factor $f(y_t,\Delta y_{t0},\Delta y_{t2})$ from Eq.~(\ref{stuff}) that raises the apparent tail at larger \mt. 
Power-law exponent $n_2 = 14$ may thus be a lower limit.

\section{2.76 $\bf TeV$ $\bf Pb$-$\bf Pb$ quadrupole spectra} \label{quadspec}

The 200 GeV quadrupole spectrum analysis reported in Ref.~\cite{quadspec}, based on MB $v_2(p_t)$ data for identified hadrons with limited statistics, established a novel analysis method. Recent $v_2(p_t,n_{ch})$ data from the LHC offer the possibility of high-statistics analysis including collision-energy and \aa\ centrality dependence of quadrupole spectra. The \auau\ analysis of MB data is here extended to cover centrality dependence of \pbpb\ data, including variation of monopole boost $\Delta y_{t0}(n_{ch})$.

\subsection{2.76 TeV Pb-Pb $\bf v_2(p_t,n_{ch})$ data}

Figure~\ref{alice1} shows published $v_2\{\rm SP\}(p_t,n_{ch})$ data for four hadron species from 15 million 2.76 TeV \pbpb\ collisions in seven centrality bins: 0-5\%, 5-10\%, 10-20\%, 20-30\%, 30-40\%, 40-50\% and 50-60\%~\cite{alicev2ptb}. The method employed for that analysis is the so-called {\em scalar-product} (SP) method~\cite{2004}.\footnote{SP is also used in this text to denote {\em single-particle} spectra.} 
Error bars represent statistical plus systematic uncertainties combined in quadrature. This conventional plotting format conceals essential information carried by $v_2(p_t,n_{ch})$ data (e.g.\ details of $\Delta y_{t0}$) that may be relevant to hydro theory. The first step in reduction of differential $v_2(p_t,n_{ch})$ data to common loci is to rescale data with measured \pt-integral $v_2(n_{ch})$ values.

\begin{figure}[h]
     \includegraphics[width=3.3in]{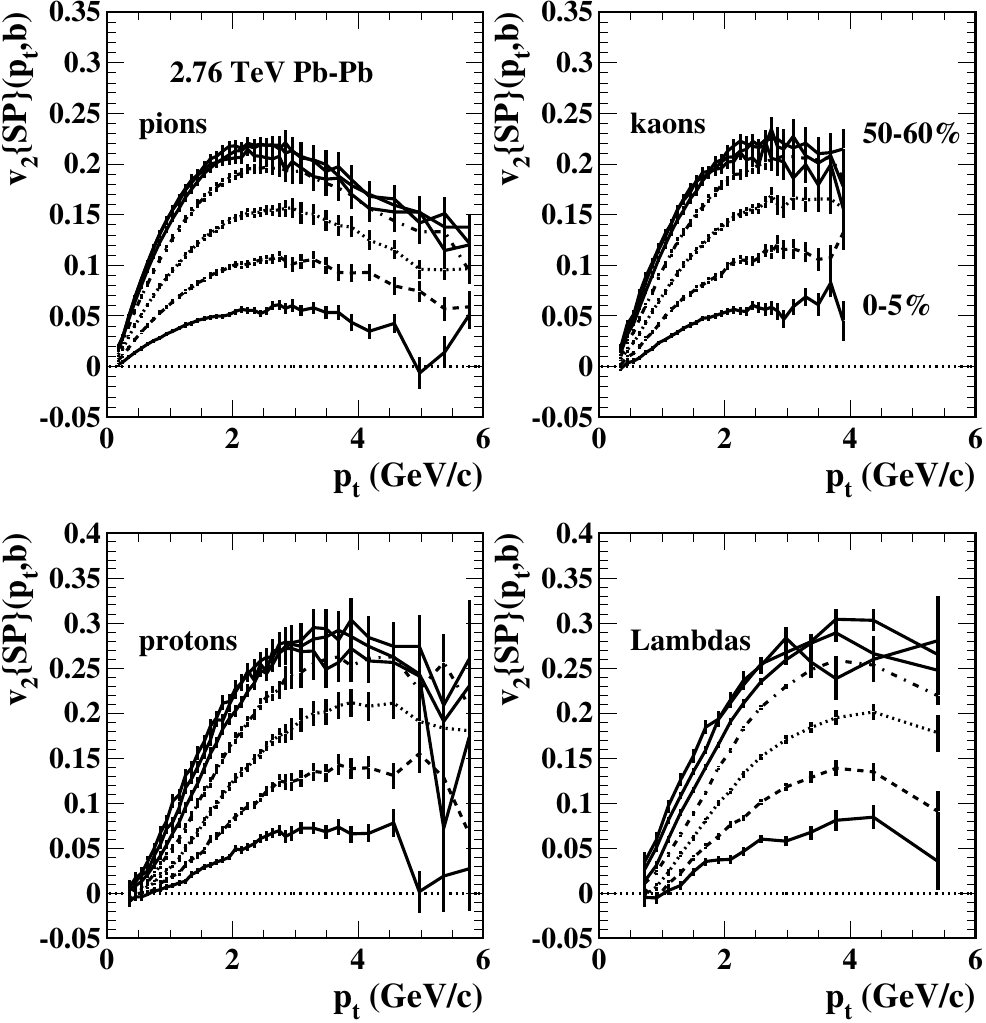}
\put(-140,233) {\bf (a)}
\put(-22,233) {\bf (b)}
\put(-140,108) {\bf (c)}
\put(-22,108) {\bf (d)}
\caption{\label{alice1}
$v_2(p_t,n_{ch})$ data for four hadron species from seven centrality classes of 2.76 TeV \pbpb\ collisions~\cite{alicev2ptb}. Bars represent statistical and systematic errors combined quadratically. Line styles proceed from most-central data as solid, dashed, dotted, dash-dotted with solid thereafter.
} 
\end{figure}

\subsection{Transforming $\bf v_2(p_t,n_{ch})$ data to common loci}

To test the extent of variation with centrality and collision energy of LHC vs RHIC $v_2(p_t,n_{ch})$ data the $v_2$ data are here rescaled by integral $v_2(n_{ch})$ values for each energy. Comparisons provide a detailed differential study. Note that within this study event classes are denoted by an associated event-class index $n_{ch}$ (within some fiducial $\eta$ acceptance) rather than a cross section $\sigma$ or impact parameter (centrality) $b$. That strategy follows from the discovery that classical Glauber Monte Carlo determination of collision geometry is questionable~\cite{ppbpid,tomglauber}.

Figure~\ref{alice4} (left) shows 2.76 TeV $v_2\{4\}(n_{ch})$ data (solid)~\cite{alicev2b} vs \pbpb\ centrality measured by fractional cross section $\sigma / \sigma_0$.
Open symbols are 62 and 200 GeV \auau\ $v_2\{\rm 2D\}(n_{ch})$ data obtained from model fits to 2D angular correlations~\cite{v2ptb}. The 62 GeV data are 10\% lower per the energy trends in Fig.~\ref{glaubtrend} (right). The dotted curve describing 200 GeV data  is derived from Eq.~(\ref{magic})~\cite{davidhq,noelliptic}
\bea \label{magic}
(2\bar \rho_0 /N_{part}) A_Q\{\rm 2D\}(n_{ch}) &=& 0.0022 N_{bin} \epsilon_{opt}^2(n_{ch}).~~~~
\eea
The 200 GeV NSD \pp\ point
is  derived from $V_2^2 = \bar \rho_0^2 v_2^2 \approx 0.006$ in Fig.~\ref{quadv2} (left) $\rightarrow v_2 \approx 0.03$ with $\bar \rho_0 \approx 2.5$.
$v_2\{4\}(n_{ch})$ data (solid) are used to rescale the $v_2\{\rm SP\}(p_t,n_{ch})$ data below that test factorization of $v_2\{\rm SP\}(p_t,n_{ch})$. 
This format may be used to compare different data sets but has the disadvantage of relying on a questionable classical Glauber model for centrality determination~\cite{tomglauber}. Figure~\ref{alice4} (right) is discussed below. 

\begin{figure}[h]
     \includegraphics[width=3.3in]{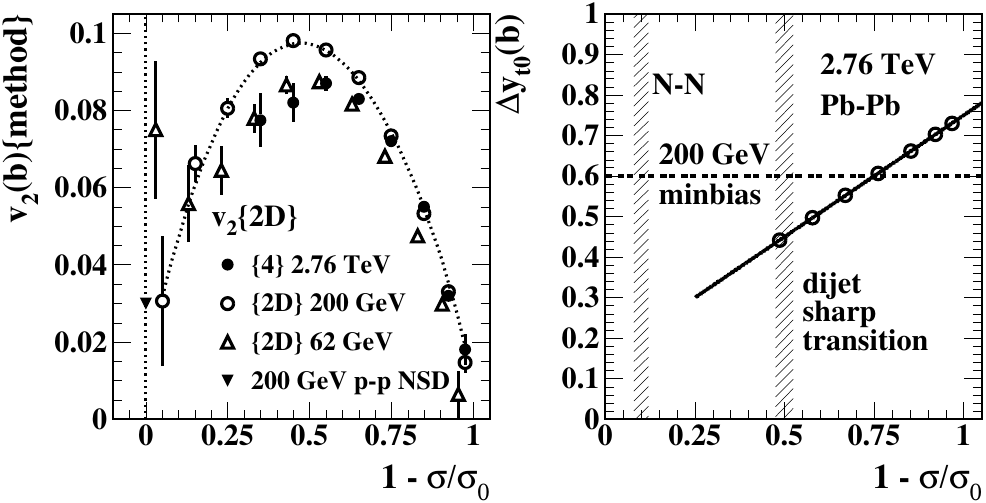}
\caption{\label{alice4}
Left: $v_2\{4\}(n_{ch})$ \pt-integral data for eight centralities of 2.76 TeV \pbpb\ collisions (solid points)~\cite{alicev2b} compared to 62 and 200 GeV $v_2\{\rm 2D\}(n_{ch})$ data trend (open points~\cite{davidhq,v2ptb}) and 200 GeV Eq.~(\ref{magic}) (dotted curve)~\cite{davidhq,noelliptic}. The 200 GeV \pp\ point is described in the text.
Right: Quadrupole source boosts $\Delta y_{t0}(n_{ch})$ for seven centralities of 2.76 TeV \pbpb\ collisions (points) inferred from $v_2(p_t,n_{ch})$ data in this study.
} 
\end{figure}

Figure~\ref{alice2} shows data from Fig.~\ref{alice1} rescaled by factor $1/v_2(n_{ch})$ where 
$v_2(n_{ch})$ is \pbpb\ $v_2\{4\}(n_{ch})$ data in Fig.~\ref{alice4} (left).
The bold dashed curves are the curves passing through 200 GeV MB $v_2(p_t)$ data in Fig.~\ref{x1} (left) derived in Ref.~\cite{quadspec} divided by 200 GeV $v_2(\text{MB}) \approx 0.055$. The solid triangles in panel (a) are derived from more-recent MB 200 GeV pion data with higher statistics~\cite{newstarpion} that agree well with  rescaled LHC data. Compare panel (a) with Fig.~\ref{x1} (left).
{  Note that the two highest curves in each panel are the two most-central event classes. As a ratio of two spectra $v_2(p_t)$ tends to have large statistical errors at higher \pt\ whereas for $v_2(p_t)/p_t$ in Fig.~\ref{alice3}} the apparent fluctuations are much reduced. And in Fig.~\ref{alice8c} (where single-particle spectra have been removed from ratio $v_2(p_t)$) the error bars are smaller than the points over the entire \pt\ range.}

\begin{figure}[h]
     \includegraphics[width=3.3in]{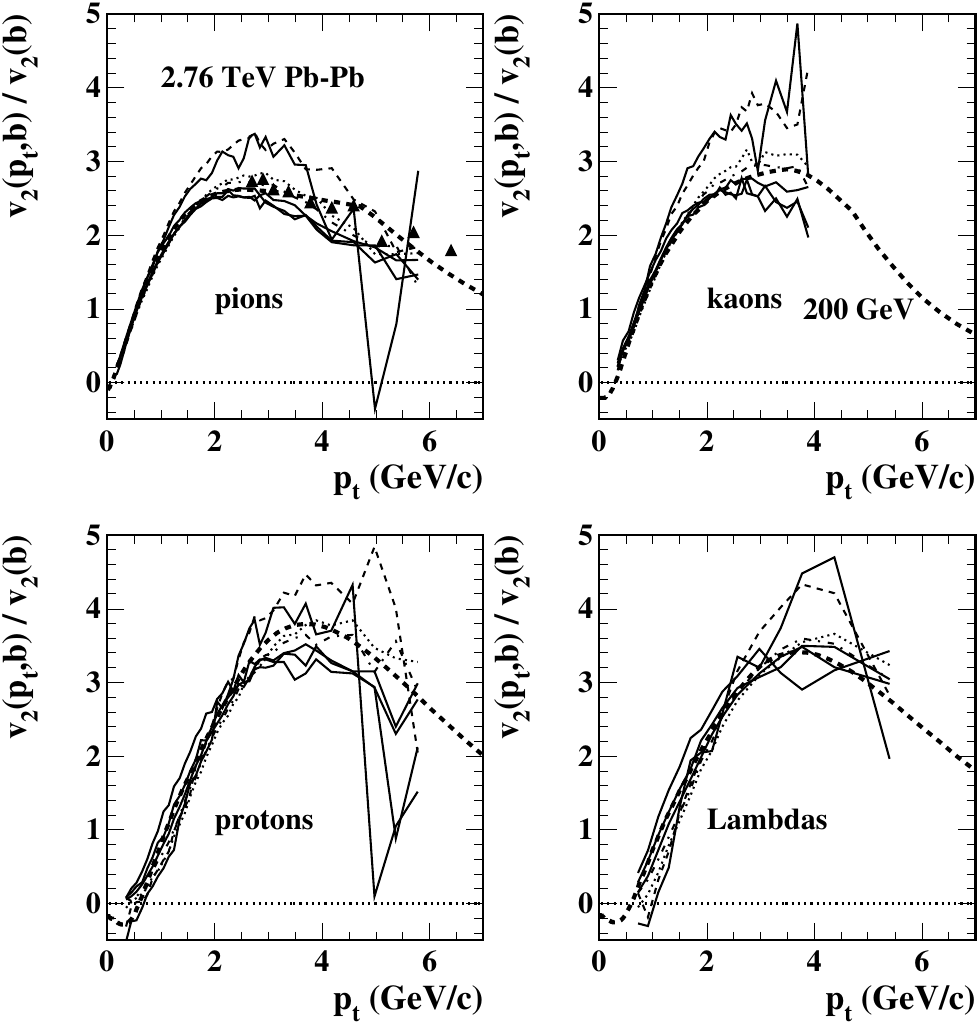}
\put(-140,233) {\bf (a)}
\put(-22,233) {\bf (b)}
\put(-140,108) {\bf (c)}
\put(-22,108) {\bf (d)}
\caption{\label{alice2}
Data from Fig.~\ref{alice1} rescaled by factor $1 / v_2\{4\}(n_{ch})$ derived from 2.76 TeV \pbpb\ $v_2(n_{ch})$ data in Fig.~\ref{alice4} (left). Bold dashed curves here are the same that appear in Fig.~\ref{x1} (left) (three line styles there for 200 GeV MB data~\cite{quadspec} scaled by factor 1/0.055. Solid triangles in panel (a) are recent high-statistics 200 GeV pion data from Ref.~\cite{newstarpion} that agree with rescaled LHC data. High curves in each panel are the two most-central event classes. Similarly for Figs.~7 and 8.
} 
\end{figure}

\begin{figure}[b]
     \includegraphics[width=3.3in]{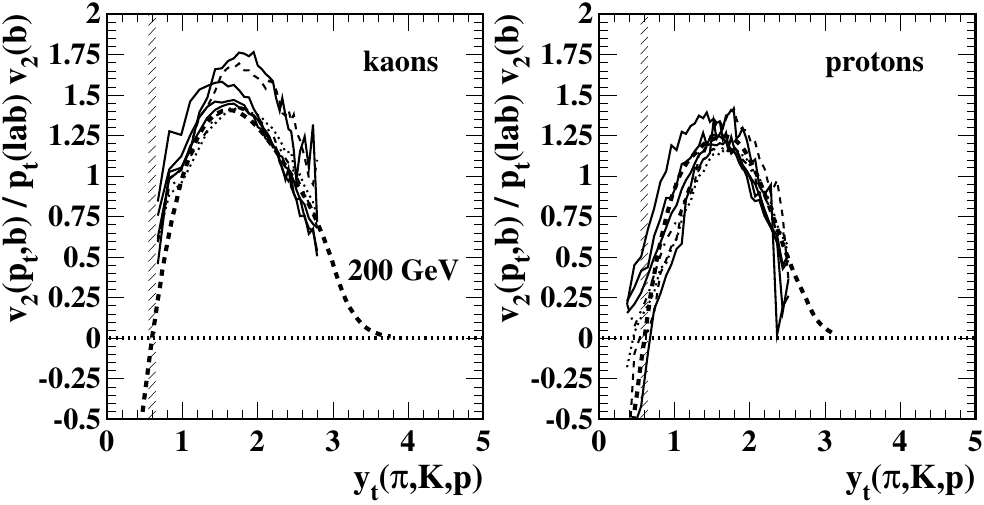}
 \caption{\label{alice3shift}
Rescaled $v_2(p_t,n_{ch})$ data from Fig.~\ref{alice2} divided by $p_t$ in the lab frame for two hadron species. A systematic variation in the apparent source boost (zero intercept) is apparent. The hatched bands indicate the nominal monopole boost $\Delta y_{t0} = 0.6$ inferred in Ref.~\cite{quadspec} from 200 GeV \auau\ MB $v_2$ data.
} 
\end{figure}

Figure~\ref{alice3shift} shows data for kaons and protons from Fig.~\ref{alice2} divided by \pt\ in the lab frame and plotted on proper \yt\ for each hadron species as in Fig.~\ref{x1} (right). The bold dashed curves in this figure correspond to those in Fig.~\ref{x1} (right), again divided by $v_2(\text{MB}) = 0.055$. The general trend is a zero intercept near $y_t = 0.6$ (dashed) as in the 200 GeV MB study, interpreted as a boost common to several hadron species. However, close examination of those data reveals systematic variation of source boost $\Delta y_{t0}$ (i.e.\ \yt\ zero intercept) with event-class index $n_{ch}$.

Figure~\ref{alice4} (right) shows boost deviations (from 0.6) required to bring data as in Fig.~\ref{alice3shift} onto a common locus corresponding to source boost $\Delta y_{t0} = 0.6$. 

Figure~\ref{alice3} shows data from Fig.~\ref{alice2} divided by \pt\ in the lab frame and plotted on proper \yt\ for each hadron species as in Fig.~\ref{x1} (right). In this case the 2.76 TeV data for seven centralities are shifted on \yt\ according to Fig.~\ref{alice4} (right), thereby emulating a common fixed boost $\Delta y_{t0} = 0.6$. The data for each of four hadron species coincide over seven \pbpb\ centralities within their uncertainties and with equivalently-scaled 200 GeV MB trends (dashed). 
Those results demonstrate that within point-to-point uncertainties ratio $v_2(p_t,n_{ch}) / v_2(n_{ch})$ is the same {\em in shape and magnitude} at 200 GeV and 2.76 TeV. {   In panel (c) the solid curve S  represents a viscous hydro prediction~\cite{shen}.} {  Also in panel (c) strong negative excursions below $y_t \approx 0.6$ may provide information about a monopole boost distribution. See Sec. \ref{monboost} and Fig.~\ref{booststudy}.} 

\begin{figure}[h]
    \includegraphics[width=3.3in]{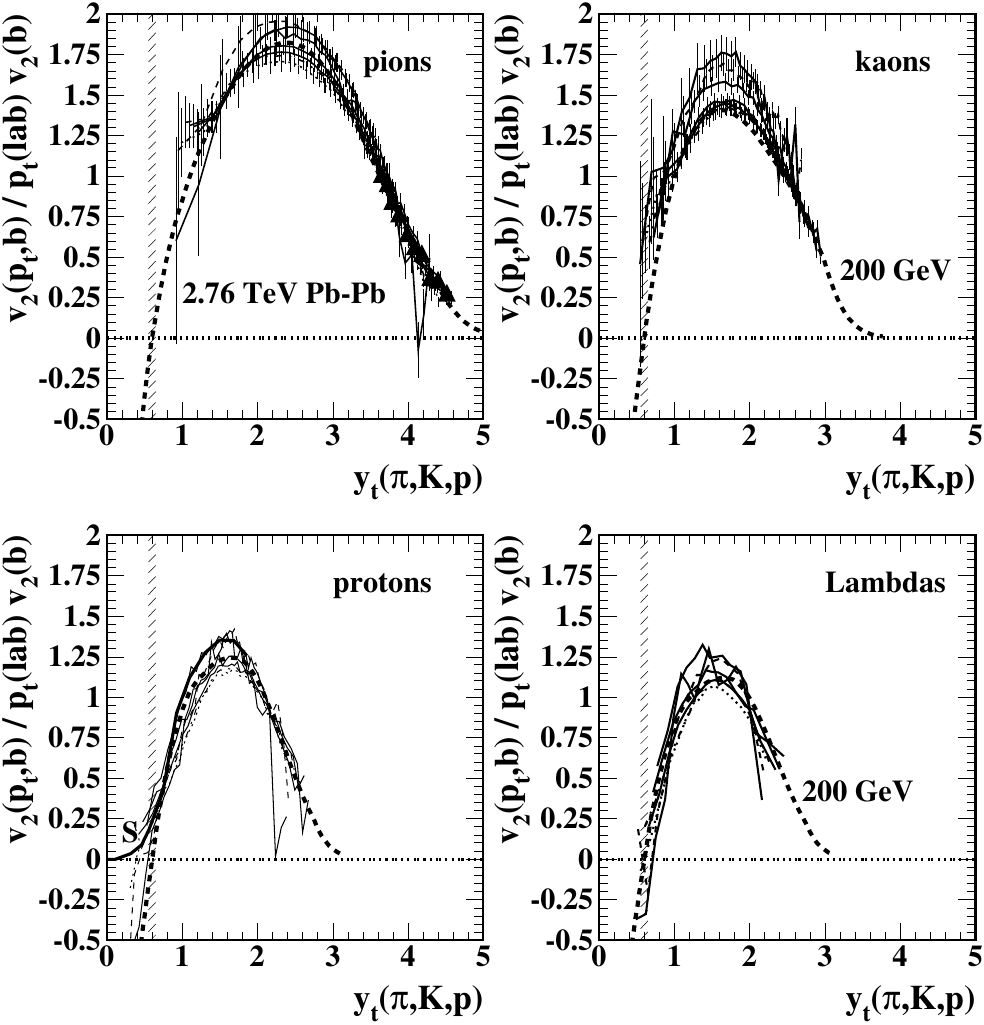}
 \put(-140,222) {\bf (a)}
\put(-22,222) {\bf (b)}
\put(-140,95) {\bf (c)}
\put(-22,95) {\bf (d)}
\caption{\label{alice3}
Data prepared as in Fig.~\ref{alice3shift} but shifted on \yt\ to a common source boost $\Delta y_{t0} = 0.6$ based on the boost trend in Fig.~\ref{alice4} (right). The dashed curves are  from Fig.~\ref{x1} (right).  The solid triangles in panel (a) are more-recent high-statistics 200 GeV pion data from Ref.~\cite{newstarpion}. Error bars are included for data in (a,b) but would obliterate any detail for data in (c,d). {   Solid curve S in panel (c) is a viscous-hydro prediction reported in Ref.~\cite{shen} for protons from 10-20\% central 2.76 TeV \pbpb\ .}
} 
\end{figure}

\subsection{SP spectra vs quadrupole spectra}

The next step in deriving quadrupole spectra requires SP spectra for identified hadrons compatible with  $v_2(p_t,n_{ch})$ data. PID spectra for 2.76 TeV \pbpb\ collisions are obtained from Ref.~\cite{alicepbpbpidspec}.
Since data trends (shape variations) in Fig.~\ref{alice3} do not change significantly with centrality, $v_2$ data for 30-40\% central are adopted as representative. That choice minimizes possible jet (nonflow) contributions relative to NJ quadrupole and facilitates spectrum modeling to obtain correspondence with \pt\ values for $v_2(p_t,n_{ch})$ data as in Fig.~\ref{alice5} of App.~\ref{spspec}.

\begin{figure}[h]
     \includegraphics[width=3.3in]{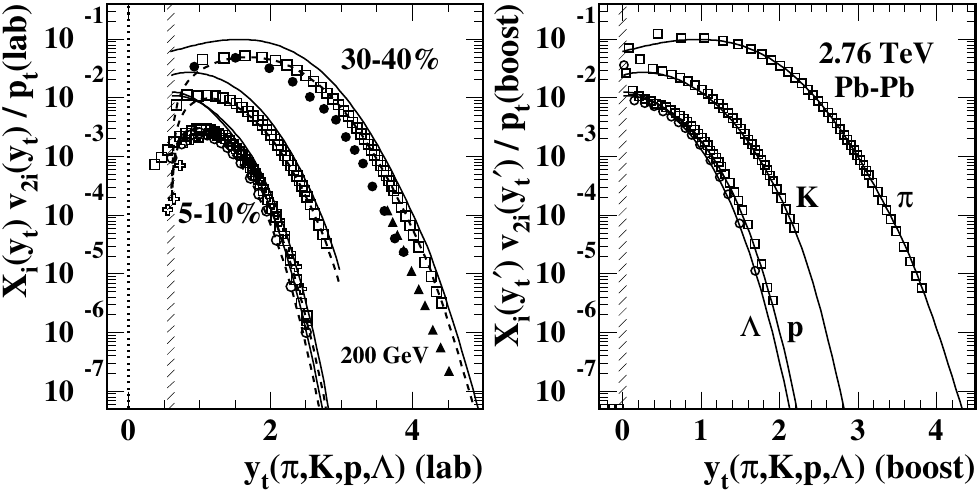}
\caption{\label{alice8ab}
Left: 2.76 TeV data and 200 GeV dashed curves from Fig.~\ref{alice3} multiplied by corresponding SP spectra in the form $X_i(y_t)$ for the 30-40\% centrality class of 2.76 TeV \pbpb\ collisions (rescaled as in App.~\ref{spspec}). {The crosses are proton data for 5-10\% central \pbpb.}
The solid dots are 200 GeV pion data from Fig.~\ref{x2} (right) rescaled to match \pbpb\ data at lower \yt.
Right: \pbpb\ data and dashed curves from the left panel divided by ratio $p_t' / p_t$ derived from Eq.~(\ref{kine}) and shifted by $\Delta y_{t0} = 0.6$ to the boost frame. The transformed dashed curves become the solid curves which are repeated also in the left panel for comparison (but in the lab frame).
}  
\end{figure}

Figure~\ref{alice8ab} (left) shows data for four hadron species in Fig.~\ref{alice3} multiplied by $v_2(n_{ch})$ (to remove the previous rescale) and by corresponding SP spectra $\bar \rho_{0i}$ as $\text{X}_i(y_t)$ (see App.~\ref{spspec}).  The data are then in a form proportional to Eq.~(\ref{stuff}).
{The crosses are additional 5-10\% central proton data to demonstrate low-\yt\ behavior for an event class proximate to $\Delta y_{t0} \approx 0.6$. See Fig.~\ref{booststudy} (right).}
Curves through $v_2(n_{ch})/p_t$ data in Fig.~\ref{x1} (right) are processed similarly to obtain several dashed curves through 2.76 TeV data in this panel. 
Solid curves are copied from the right panel for comparison. 
Solid points are 200 GeV pion data from Fig.~\ref{x2} (right) rescaled to match higher-energy data at lower \yt\ for comparison. This panel corresponds to Fig.~9 of Ref.~\cite{quadspec}.

Figure~\ref{alice8ab} (right) shows data from the left panel (lab frame) divided by factor $p'_t / p_t$ from Eq.~(\ref{kine}) and transformed  to the boost frame (shifted left by $\Delta y_{t0}$ = 0.6) to obtain data  proportional to quadrupole spectra $\bar \rho_{2i}(y_{ti}',n_{ch})$ for four hadron species. The solid curves here are dashed curves from the left panel treated the same. The last step is transformation to $m_{ti}' - m_i$.

\begin{figure}[t]
     \includegraphics[width=3.3in]{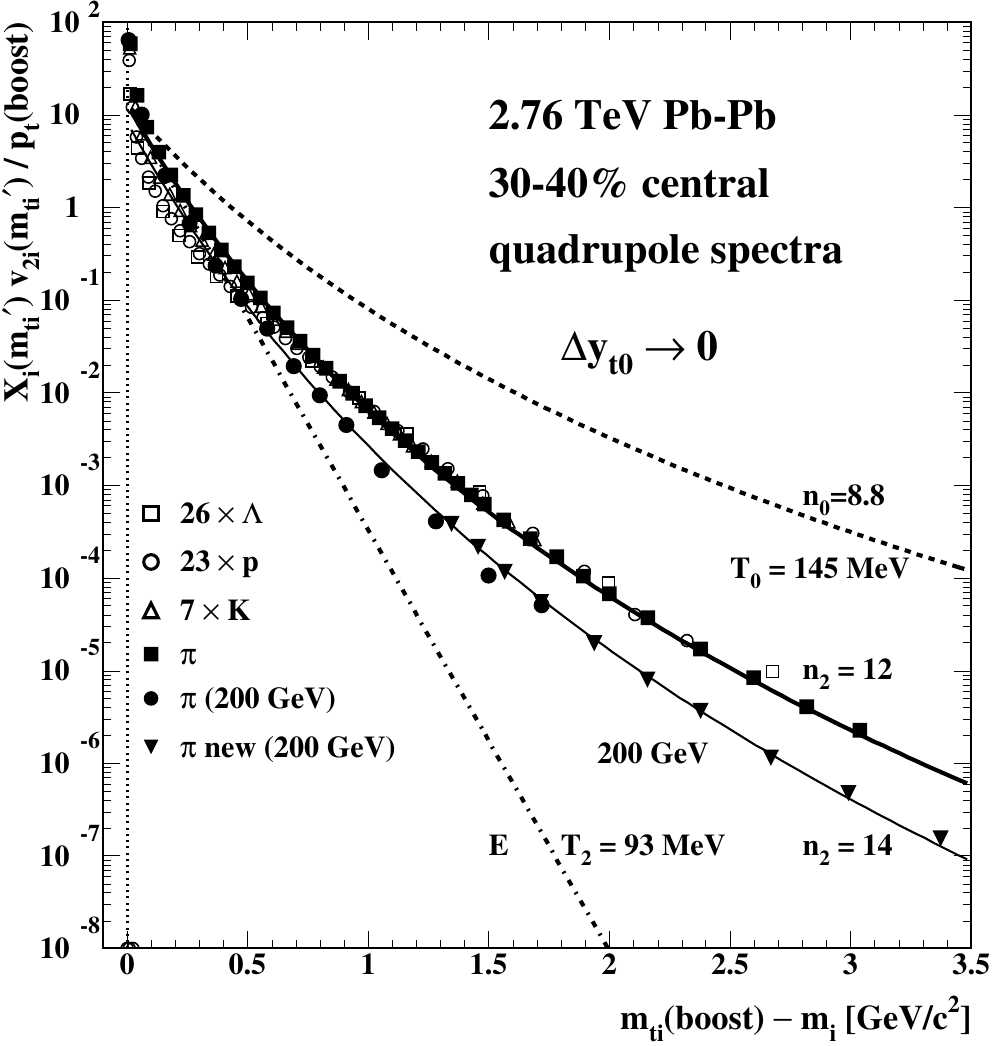}
\put(-80,145) {$\bf \hat S_0(m_t)$}
\put(-99,120) {$\bf \hat S_2(m_t)$}
\caption{\label{alice8c}
Data from Fig.~\ref{alice8ab} (right) transformed  to $m_{ti}' - m_i$ and rescaled (relative to pions) as indicated in the plot. Error bars for pion data are increased 3-fold for visibility. The bold solid curve is model distribution  $\hat S_2(m'_t)$ that describes rescaled 2.76 TeV data. Dash-dotted curve E is the Boltzmann-exponential limit for $T_2 = 94$ MeV. The dashed curve is proportional to SP soft component $\hat S_0(m_t)$ for unidentified hadrons from 2.76 TeV \pp\ collisions derived in Ref.~\cite{alicetomspec}. The thin solid curve, solid squares and inverted solid triangles are the 200 GeV quadrupole spectrum from Fig.~\ref{xbig} rescaled to match 2.76 TeV data at low \mt\ for shape comparison.
} 
\end{figure}

Figure~\ref{alice8c} shows quadrupole spectra for four hadron species transformed to densities on $m_{ti}' - m_i$ and rescaled as noted on the plot (corresponding to  200 GeV results in Fig.~\ref{xbig} from Ref.~\cite{quadspec}). Above 0.5 GeV/$c^2$ the spectra coincide as for 200 GeV data but below that point there are significant deviations. 
Otherwise,  quadrupole spectra for four hadron species at 2.76 TeV are well described by a single model function $\hat S_2(m'_t)$ (bold solid curve) with $T_2 \approx 93$ MeV and $n_2 \approx 12$.  The dash-dotted curve is the Boltzmann-exponential E equivalent with $1/n_2 \rightarrow 0$. Those results may be compared with  $T_2 \approx 93$ MeV and $n_2 \approx 14$ for 200 GeV pion data from Fig.~\ref{xbig} (solid dots, inverted triangles and thin solid curve) rescaled to match 2.76 TeV results at lower \mt.
The dashed curve is proportional to pion SP  soft component $\hat S_0(m_t)$ for 2.76 TeV \pp\ collisions~\cite{alicetomspec} plotted in the boost frame for comparison. The quadrupole spectrum is clearly quite  distinct from the SP-spectrum $\hat S_0$ soft component.

The data-derived quantity in Fig.~\ref{alice8c} is proportional to
\bea \label{stuff2}
\frac{V_{2i}(m_t',n_{ch})}{p'_t} &=&  f(m_t',\Delta y_{t0},\Delta y_{t2})
\\ \nonumber 
&\times&   \left\{\frac{\Delta y_{t2}}{2T_2}\right\}\, \bar \rho_{2i}(m_t',b,T_2,n_2)
\\ \nonumber 
&\approx&   \left\{\frac{\Delta y_{t2}}{2T_2}\right\} \bar \rho_{2i}(n_{ch})\, \hat S_2(m_t',T_2,n_2)
\eea
plotted as points for four hadron species. The function $f(m_t',\Delta y_{t0},\Delta y_{t2})$ is unity at lower \mt\ but increases monotonically with increasing \mt\ at a rate determined by the unknown ratio $\Delta y_{t2}/\Delta y_{t0}$. Exponent $n_2 \approx 12$ is then a lower limit for the actual quadrupole spectrum. Unit-normal $\hat S_2(m_t',T_2,n_2)$ estimates the functional form of a universal quadrupole spectrum shape for 2.76 TeV. 

Product $\Delta y_{t2}\bar \rho_{2i}(n_{ch})$ represents the ``amplitude'' of the NJ quadrupole. In turn, $\bar \rho_{2i} = z_{2i} \bar \rho_2$ expresses a relation similar to that for SP spectrum integrals $\bar \rho_{0i} = z_{0i} \bar \rho_0$, where fractional abundances $z_{0i}$ may be consistent with the statistical model~\cite{statmodel}. Figs.~\ref{xbig} and \ref{alice8c} suggest that $z_{2i} \approx z_{0i}$ At present there is no clear way to determine factors $\Delta y_{t2}$ and $\bar \rho_s(n_{ch})$ separately but  limiting cases may be considered. The condition $\Delta y_{t2} < \Delta y_{t0}$ (positive-definite boost) determines a lower limit on quadrupole density $\bar \rho_{2}$. Comparison of the distinctive shape of the NJ quadrupole spectrum (cutoff at $\Delta y_{t0}$ and very soft spectrum) with SP spectra may establish an upper limit on $\bar \rho_{2}$. In Ref.~\cite{quadspec} an upper limit on pion $\bar \rho_{2}$ of 5\% of  $\bar \rho_0$ was estimated by such a comparison. Section~\ref{hydropredict} presents some additional criteria for estimation.

\subsection{Mesons, baryons and NCQ Scaling} \label{ncqstuff}

In Fig.~5 of Ref.~\cite{alicev2ptb} $v_2(p_t)$ vs \pt\ data are shown for eight hadron species plotted together for each of seven \pbpb\ centralities. Concerning that figure ``A clear mass ordering is seen for all centralities in the low-\pt\ region (i.e. $p_t \leq$ 3 GeV/c), attributed to the interplay between elliptic and radial flow. For higher values of \pt\ (i.e.\ $p_t > 3$ GeV/c) particles tend to group [vertically] according to their type, i.e.\ mesons and baryons.'' Subsequently, two alternative plotting formats are considered as forms of ``scaling'' in which $x$ and $y$ variables are multiplied by factors apparently intended to test various hypotheses.

In its Fig.~8 data are replotted after ``NCQ [number of constituent quarks] scaling'' wherein $x$ and $y$ variables are each divided by a number of quarks $n_q$: 2 for mesons and 3 for baryons. This has the effect of reducing displacements on \pt\ at lower \pt\ associated with ``mass ordering'' and  the vertical differences between mesons and baryons at higher \pt. The result is conventionally seen as confirming a model in which quarks dominate collision dynamics earlier in the collision, and hadronization proceeds by quark coalescence~\cite{rudy,duke,tamu}. However, in its Fig.~9 ratios of scaled data to proton data show large deviations above and below unity depending on hadron species.

In its Fig.~10 data are replotted according to \mt\ scaling. That is, the $x$ axis is transformed from $p_t/n_q$ to $(m_{ti} - m_i) / n_q$. Displacements on the $x$ axis at lower \pt\ are further reduced from those in Fig.~8 but  vertical displacements at higher \pt\ remain unchanged. Its Fig.~11 shows that large deviations persist among hadron species.

\begin{figure}[h]
     \includegraphics[width=3.3in]{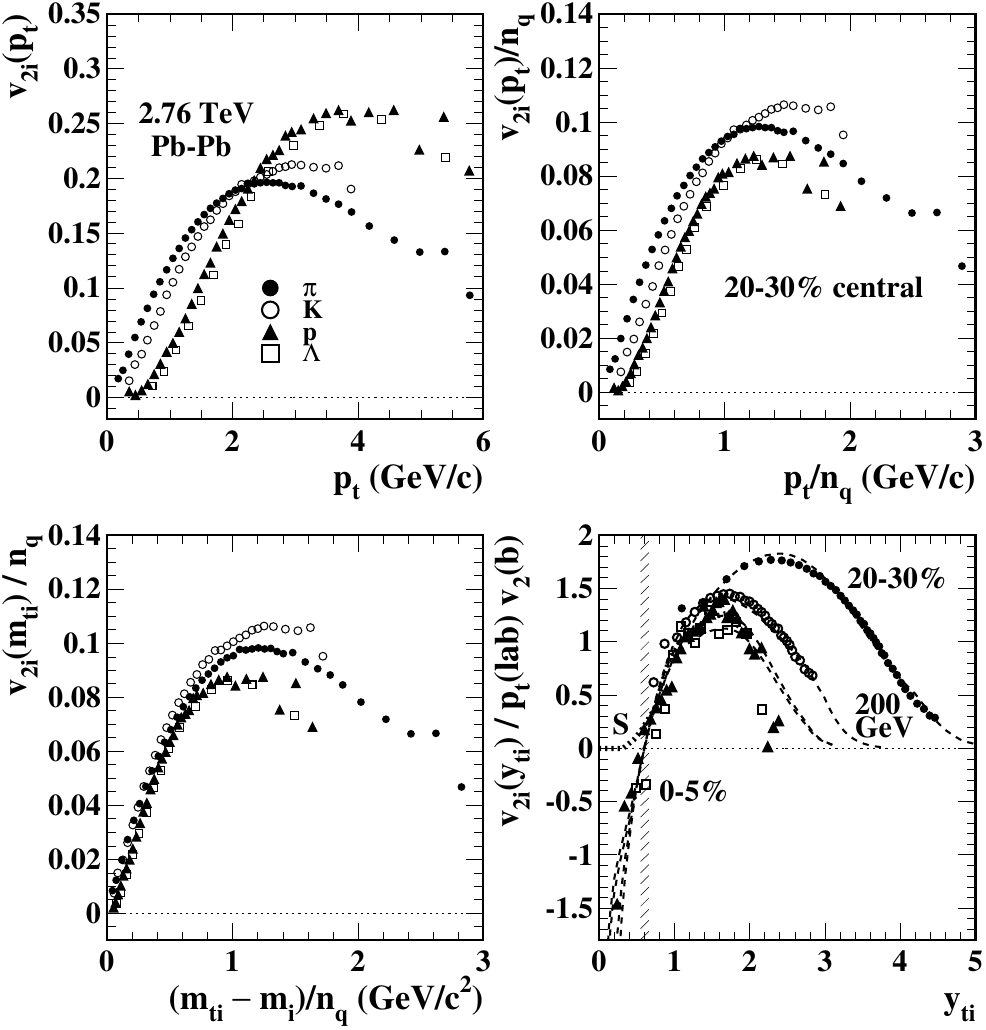}
\put(-140,233) {\bf (a)}
\put(-22,233) {\bf (b)}
\put(-140,108) {\bf (c)}
\put(-22,95) {\bf (d)}
\caption{\label{alice1xx}
(a) $v_{2}(p_t)$ data vs \pt\ for pions, charged kaons, protons and Lambdas from 20-30\% central 2.76 TeV \pbpb\ collisions plotted in a conventional format. 
(b) The same data plotted according to so-called NCQ or constituent-quark scaling.
(c) The same data plotted vs ``kinetic energy'' or transverse mass $m_{ti} - m_i$ for hadron species $i$.
(d) $v_2(p_t)$ data plotted in a format corresponding to Eq.~(\ref{combfac}) {per Fig.~\ref{alice3}. Proton and Lambda data here are 0-5\% central to emphasize their negative trends at lower \yt. Different centralities are shifted to common $\Delta y_{t0} = 0.6$ per Fig.~\ref{alice4} (right). Curve S is hydro theory for protons from Ref.~\cite{shen} with upper limit near \yt\ = 1.9.}
The 200 GeV dashed curves are predictions taken from Fig.~\ref{x1} (right).
} 
\end{figure}

Figure~\ref{alice1xx} below considers Ref.~\cite{alicev2ptb} scaling exercises in the context of the present study. The symbol sizes are reduced to make small differences visible. Panel (a) corresponds to Fig.~\ref{x1} (left) which has been described in detail in Sec.~\ref{200gevquad}. In that section data trends in the conventional format are simply related  to a single {\em universal} functional form (quadrupole spectrum) representing a common boosted hadron source for all hadron species.

Panel (b) shows that NCQ scaling reduces horizontal displacements on \pt\ at lower \pt\ and likewise vertical displacements at higher \pt\ between baryons and mesons, which was the {\em desired effect} motivating that  scaling. But given the context provided by the present study that maneuver has no physical basis and is therefore misleading.

Panel (c) illustrates \mt\ scaling wherein the separations between zero intercepts at lower \pt\ (``mass ordering'') corresponding to monopole boost $\Delta y_{t0}$ are further reduced, but that is also a misleading result. Because \pt\ relates to \yt\ via $p_t = m_i \sinh(y_{ti})$, to good approximation at lower \pt\ the  intercepts in (a) are located near $m_i \Delta y_{t0}$ and those in (b) near $(m_i/n_{q_i}) \Delta y_{t0}$. However, because $m_{ti} - m_i = m_i [\cosh(y_{ti}) - 1] \approx m_i y_{ti}^2/2$  the intercepts in (c) occur near $(m_i/n_{q_i}) \Delta y_{t0}^2/2$. Since $\Delta y_{t_0} \approx 0.6$, $\Delta y_{t0}^2/2 \approx 0.18$. Thus, apparent displacements in panel (c) are reduced from (b) by a factor 3. 

Panel (d) corresponds to Fig.~\ref{x1} (right) and Fig.~\ref{alice3}. The most important result derivable from $v_2(p_t)$ data is clearly  shown here as the zero intercept near $\Delta y_{t0} \approx 0.6$. {The strong negative trend of data below that point conflicts with a viscous hydro theory prediction (bold dotted) from Ref.~\cite{shen}. Data trends at lower \pt\ and theory curve S are further discussed in Sec~\ref{monboost}.} 

Data structures at higher \pt\ are simply related to statistical-model hadron abundances and to the fact that $y_{ti} \approx \ln(2p_t / m_i)$ at higher \pt. Figure~\ref{alice8c} demonstrates that in an optimized plot format the several spectra are {\em identical in shape}. One may conclude that the ``scalings'' in (b) and (c) {\em act to obscure} to varying degrees  the intercept at $\Delta y_{t0} \approx 0.6$. The results in (a) produce confusion because of the poor properties of $v_2(p_t)$. Response to that confusion has been to introduce {\em ad hoc} procedures that {\em seem} to simplify data trends but instead discard the most important information carried by data. 

Results in panels (b,c) may be contrasted with overall results for quadrupole spectrum analysis as in the present study: $v_2(p_t)$ vs \pt\ data in Figs.~\ref{x1} (left) and \ref{alice1} are transformed simply to quadrupole spectra, {\em all with a single common form} in the boost frame in Figs.~\ref{xbig} and \ref{alice8c} with accurate determination of spectrum parameters, whereas (b,c) represent {\em ad hoc} attempts to accomplish a similar transformation with confusing result. Panel (d) is an intermediate step in  quadrupole spectrum analysis.

Section~\ref{quadspec} describes analysis of \pbpb\ \pt-differential $v_2(p_t)$ data to obtain corresponding quadrupole \mt\ spectra as in Fig.~\ref{alice8c} with their several properties derived for a specific  collision energy and event class. In the following Sec.~\ref{edep} \pbpb\ event-\nch\ dependence and collision-energy dependence of \pt-integral quadrupole {\em correlation amplitudes}, measured by {\em extensive} quantity $V_2^2$ representing number of correlated pairs, are determined.

\section{Centrality and energy trends} \label{edep}

This section first considers model fits to 2D angular correlations with amplitude measured by {\em number of correlated pairs} $V_2^2$ as a uniquely accurate quadrupole analysis method. 2D model fits applied to 200 GeV \pp\ angular correlations reveal a simple trend on soft-component density $\bar \rho_s$. The \pp\ trend is then generalized to \aa\ collisions as a conjecture. \pp\ and \aa\ data are found to be compatible on the same plot space $V_2^2$ vs $\bar \rho_0$ and validate the conjectured trend. A corresponding study of energy dependence comparing $v_2$ and $V_2^2$ measures from SPS to highest LHC energies reveals the negative effect of $v_2$ as a ratio of two strongly energy-dependent quantities.

\subsection{$\bf v_2$ via model fits to 2D angular correlations} \label{modelfits}

In  Refs.~\cite{anomalous,v2ptb,ppquad} quadrupole amplitudes were extracted via model fits to 2D {\em angular} autocorrelations on $(\eta_\Delta,\phi_\Delta)$ ($x_{\Delta}$ is a {\em difference variable} for space variable $\eta$ or $\phi$ as opposed to ``lag'' variable $\tau$ for time $t$). In Refs.~\cite{anomalous,v2ptb} a 2D charge density at midrapidity is defined by $\bar \rho_0 \equiv n_{ch}/2\pi \Delta \eta \approx d^2n_{ch}/d\phi d\eta$. The quadrupole amplitude is determined by the {\em correlated pairs per final-state hadron} represented by model parameter $A_{Q}$ (accompanied by a factor 2 in the model as indicated in Eq.~(11) of Ref.~\cite{v2ptb}). Those conditions apply for data reviewed in Sec.~\ref{200gevquad} to maintain consistency with previous publications. In more-recent analysis $\bar \rho_0 \rightarrow n_{ch}/\Delta \eta \approx dn_{ch}/d\eta$ (not a density on azimuth), and a modified model including $A_Q$ without factor 2 returns conventional values for $v_2$ when $A_Q \equiv \bar \rho_0 v_2^2$. That implies $A_Q$ data from those previous studies require a factor $2 \cdot 2\pi$ and $v_2$ data require a factor $\sqrt{2}$ to compare with LHC data. Data from those cited references are updated accordingly where used in the present study.

\begin{figure}[h]
	\includegraphics[width=1.49in,height=1.44in]{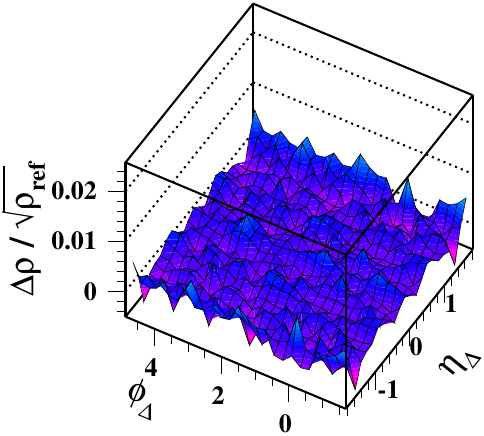}
\put(-100,85) {\bf (a)}
	\includegraphics[width=1.49in,height=1.44in]{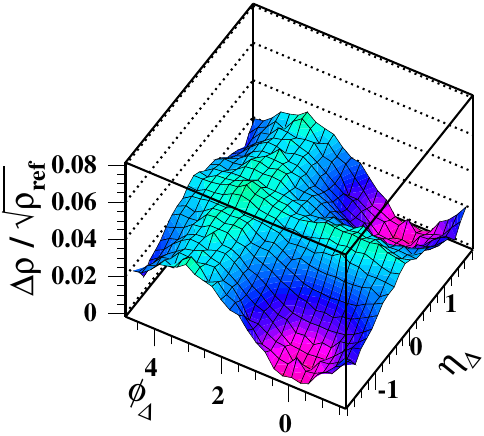}
\put(-100,85) {\bf (b)}\\
	\includegraphics[width=1.49in,height=1.44in]{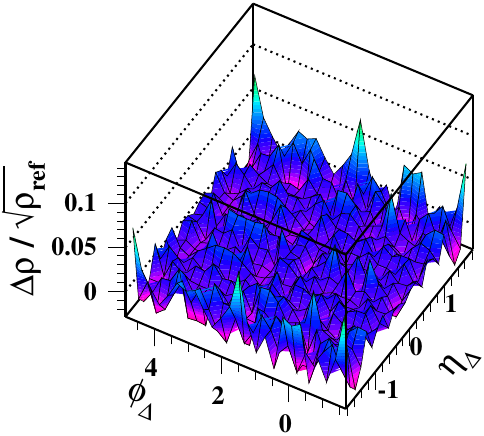}
\put(-100,85) {\bf (c)}
	\includegraphics[width=1.49in,height=1.44in]{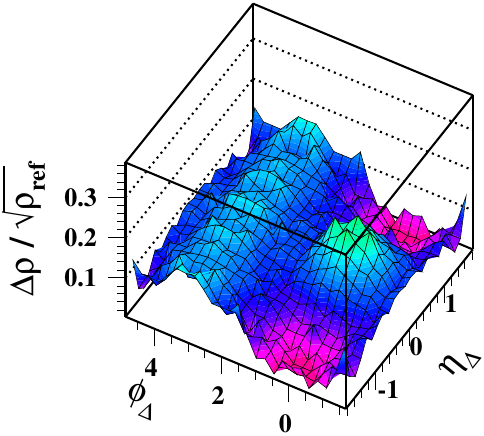}
\put(-100,85) {\bf (d)}
\caption{\label{fits} 
(Color online) Perspective views of \yt-integral 2D angular correlations as $\Delta\rho/\sqrt{\rho_{ref}} \leftrightarrow A_X$ on  $(\eta_{\Delta},\phi_{\Delta})$  from  \pp\ collisions at $\sqrt{s}$ = 200~GeV for $n=1,~6$ multiplicity classes (upper and lower rows respectively)~\cite{ppquad}. (a,c) fit residuals (vertical sensitivity increased two-fold), (b,d) jet + NJ quadrupole contributions obtained by subtracting fitted offset, soft-component and BEC/electron elements of the 2D fit model from the data histograms. 
}  
\end{figure}

Fig.~\ref{fits} shows examples of fit decomposition and residuals using 200 GeV \pp\ correlation data as described in Ref.~\cite{ppquad} from the first (a,b) and sixth (c,d) multiplicity bins.  Each row shows fit residuals (data $-$ model) and jet-related + NJ-quadrupole structure. The last are obtained by subtracting fitted model elements soft (1D Gaussian on $\eta_{\Delta}$), BE/electrons (2D exponential at origin) and offset from data histograms leaving jet-related structure and NJ quadrupole. Fit residuals (a,c) are comparable in magnitude to statistical errors. Lack of significant nonrandom structure in the residuals suggests that the standard 2D model exhausts all information in data.

Meaningful structure in (b,d) may be interpreted as follows. The structure in (b) is dominated by jets, with a same-side 2D Gaussian (jet peak) at the origin and away-side dipole $\cos(\pi - \phi_\Delta)$. The same-side jet peak is strongly elongated in the azimuth direction. The structure in (d) includes the same jet contributions but also includes a relatively strong quadrupole $\cos(2\phi_\Delta)$ contribution (increased {\em relative to the jet structure} by factor 10) that is evident in three ways: (i) The quadrupole lobe at $\pi$ is evident by the smaller radius of curvature superposed on the away-side jet dipole, (ii) the curvature at larger $|\eta_\Delta|$ near $\phi_\Delta = 0$ has been reduced to zero as the quadrupole fills in that region, and (iii) the same-side jet peak {\em appears} to be much narrower on $\phi_\Delta$ but that is actually the same-side lobe of the quadrupole {\em superposed} on the broader jet peak. The quadrupole contribution to high-\nch\ 200 GeV \pp\ collisions is thus clearly evident if the correlation structure is understood. One may contrast that result with Ref.~\cite{cmsridge} reporting a CSM ``ridge.''

\subsection{Alternative $\bf v_2$ methods} \label{alternative}

\begin{figure}[h]
	\includegraphics[width=3.3in]{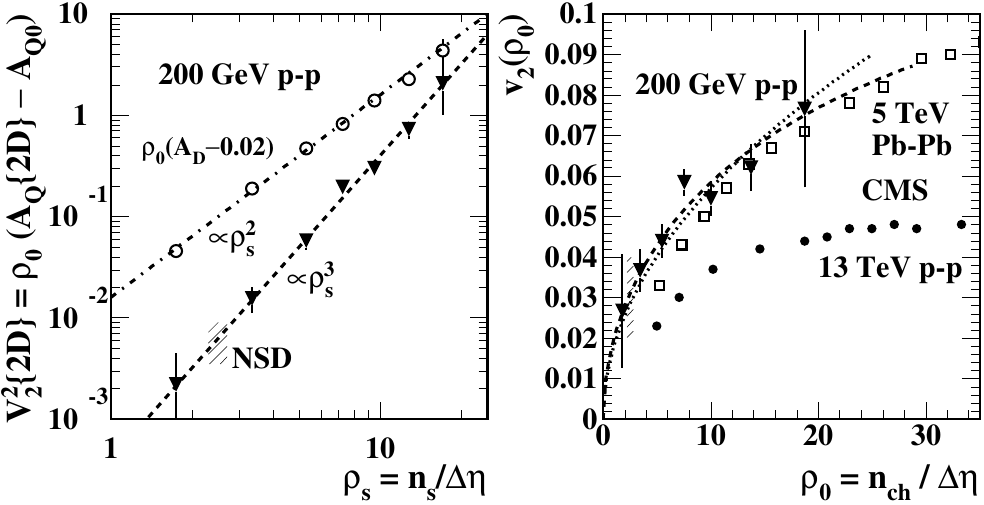}
	\caption{ \label{quadv2} 
		Left:  NJ quadrupole correlations from 200 GeV \pp\ collisions (solid triangles) measured by $V^2_2 \propto$ total number of correlated pairs vs soft-component density $\bar \rho_s$ illustrating a trend $V^2_2 \propto \bar \rho_s^3$ (dashed)~\cite{ppquad}. Also shown are away-side dipole peak amplitudes $A_D$ (open circles) treated similarly with reference trend $\propto \bar \rho_s^2$ (dash-dotted).
		Right: \pp\ data (triangles) and dashed curve from left transformed via $v_2 = \sqrt{V_2^2} / \bar \rho_0$. {13 TeV \pp\ (solid dots) and 5 TeV \pbpb\ (open squares) data  are from Fig.~1 of Ref.~\cite{cmsabv2}.}
	} 
\end{figure}

{As noted in the Introduction there is a number of ``methods'' for determining $v_2$ values from particle data by numerical methods not applied directly to {\em intact} 2D angular correlations as in Fig.~\ref{fits}. Reference~\cite{kolk} reports a comparison study of several such methods applied to 7 TeV \pp\ collisions with the goal to obtain at least an upper limit on $v_2$ for that system. In particular, $v_2$\{\text{SP}\} and $v_2$\{\text{subevent}\} methods were addressed. It was concluded that ``...nonflow [jets] is the dominant or the only correlation in 7 TeV proton-proton data at the LHC.'' Further, ``...$v_2$ in 7 TeV proton-proton collisions with at least 10 tracks is less than 0.05.'' {   The issues seem confirmed by Ref.~\cite{cmsabv2} data  in Fig.~\ref{quadv2} (right) six years later.}

{  In Sec.~\ref{quadspecmeth} conventional Eq.~(\ref{v2av})~\cite{poskvol} is based on assumed {\em event-wise} determination of an event-plane angle $\Psi_r$ from particle data. In practice that method places strong limitations on $v_2$ estimation accuracy. Model fits to 2D angular correlations have no such limits. 2D histograms  as in Fig.~\ref{fits} may be accumulated for millions of collision events without regard to a special event-wise reaction plane. Accurate results for \pp\ event classes with only a few particles per collision may thereby be obtained.}

{  Reference~\cite{poskvol} acknowledges the possibility of such ``pair-wise azimuth correlations'' relating to its Eq.~(36). However, it notes that the ``signal'' $\sim v_2^2$ may be small and warns that ``reconstruction of the triple differential distribution with respect to the reaction plane (the goal of the flow analysis) becomes more involved.'' But there is no such complexity in practice, especially as the relevance of a reaction plane is highly doubtful per the present study and Ref.~\cite{njquad}. Also, extension of the two-particle density concept from 1D azimuth alone to 2D pair distributions on $(\eta_\Delta,\phi_\Delta)$ (see Fig.~\ref{fits}) makes possible accurate separation of NJ quadrupole and other ``nonflow'' contributions {\em not possible} with a 1D approach.}

\subsection{$\bf V_2^2$ p-p quadrupole vs dipole (dijet) trends} \label{ppquadjet}

Figure~\ref{quadv2} (left) shows fitted NJ quadrupole amplitude $A_\text Q\{\rm 2D\}$ for 200 GeV \pp\ collisions in the form $V_2^2\{\rm 2D\} \equiv \bar \rho_0 A_Q \{\rm 2D\}$ (solid triangles) vs soft charge density $\bar \rho_s$~\cite{ppquad}. A small offset $A_\text{Q0}$ independent of $\bar \rho_s$ representing transverse-momentum conservation is subtracted from $A_\text Q$ data.  The dashed reference line represents {\em number of correlated pairs}  $V_2^2\{\rm 2D\} \propto \bar \rho_s^3$. The \pp\ NJ quadrupole amplitude thus {\em increases rapidly} with increasing charge multiplicity. 
{Also shown are measured amplitudes $A_\text D$ of the away-side dipole jet peak plotted as $\bar \rho_0(A_\text D- A_\text{D0})$ (open squares) for direct comparison with quadrupole $A_\text Q$ data. The dash-dotted reference line confirms the trend $\bar \rho_h \propto \bar \rho_s^2$ first reported in Ref.~\cite{ppprd}.}

Given that measured dijet (dipole) production trend one may consider its implications  within the sequence
\bea \label{ppconject}
V_2^2(n_s) \equiv \bar \rho_0^2(n_s)v_2^2(n_s)\propto \bar \rho_s^3 \rightarrow \bar \rho_s \times \bar \rho_h,
\eea
where $\bar \rho_s$ plays the role of participant (low-$x$ gluon) number $N_{part}$ and $\bar \rho_s^2 \propto \bar \rho_h$ plays the role of gluon-gluon binary-collision number $N_{bin}$~\cite{tomalicempt,alicetomspec,ppquad}.  \pp\ quadrupole data appear to be in that sense consistent with $V_2^2\{\rm 2D\} \propto N_{part} \times N_{bin}$. 
However, that phenomenological conjecture is not required by the \pp\ result. The question remains whether it is relevant for \aa\ collisions.}

{Figure~\ref{quadv2} (right) shows corresponding $v_2(\bar \rho_0)$ values inferred as $v_2 = \sqrt{V_2^2} / \bar \rho_0$, with $\bar \rho_0 = \bar \rho_s + \alpha \bar \rho_s^2$ and $\alpha \approx 0.006$ for 200 GeV collisions~\cite{ppprd}. The dashed curve at right is the dashed line at left transformed just as for the $V_2^2$ data. It then follows the trend \bea
v_2(\bar \rho_0) &\propto& \bar \rho_s^{3/2} / (\bar \rho_s + \alpha \bar \rho_s^2) 
\rightarrow \bar \rho_s^{1/2} / (1 + \alpha \bar \rho_s).
\eea
The dotted curve is $v_2 \propto \sqrt{\bar \rho_0}$ for comparison. This figure demonstrates that model fits to 2D angular correlations are capable of accurate determination of $v_2$ values (via $V_2^2$ values) for small systems and very low \nch. The quadrupole trend on $\bar \rho_0$ can be extrapolated down to zero event multiplicity. One should note the simplicity of the $V_2^2$ trend on $\bar \rho_s$ vs the complexity of the $v_2$ trend on $\bar \rho_0$ and reduction of a correlation trend $\propto \bar \rho_s^3$ to a $v_2$ trend $\propto \bar \rho_0^{1/2}$ with complete loss of physical insight.}

{ Also included at right are $v_2(n_{ch})$ data for 13 TeV \pp\ collisions (solid dots) and 5 TeV \pbpb\ collisions (open squares) from Ref.~\cite{cmsabv2} Fig.~1 compared to 200 GeV \pp\ data. The 5 TeV \pbpb\ data are well described by a \pp\  dashed curve derived from the 200 GeV \pp\ trend in the left panel but the 13 TeV \pp\ data are nearly 50\% low.}

\subsection{Comprehensive $\bf V_2^2$ A-B description} \label{multcent}

In what follows ALICE \pbpb\ $v_2(n_{ch})$ data are from Ref.~\cite{alicev2b},
ALICE $\bar \rho_0(n_{ch})$ data are from Ref.~\cite{alicerho0},
STAR \auau\  $\bar \rho_0(n_{ch})$ data are from Tables~III and IV of Ref.~\cite{anomalous} and
STAR \auau\ $v_2(n_{ch})$ and $A_\text{Q}(n_{ch})$ data are from Ref.~\cite{v2ptb}. 200 GeV \pp\ data are from Ref.~\cite{ppquad}.

\begin{figure}[h]
	\includegraphics[width=3.3in]{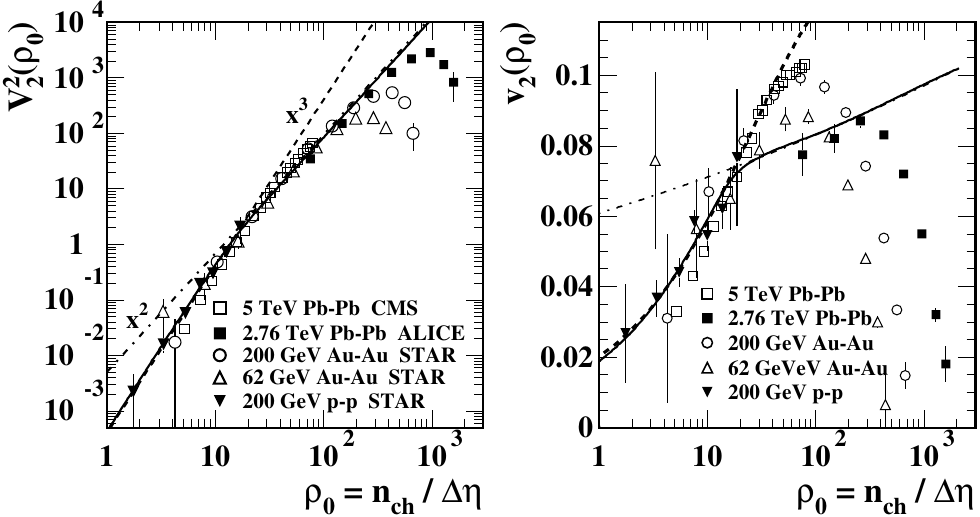}
	\caption{ \label{quadxx} 
		Left: NJ quadrupole correlations from 62 and 200 GeV \auau\ collisions (open circle and triangle) and 2.76 TeV \pbpb\ collisions (solid squares) as  $V^2_2$ vs charge density $\bar \rho_0$. {   Also included are 5 TeV \pbpb\ $v_2(p_t)$ data (open squares) from Ref.~\cite{cmsabv2}.}  200 GeV \pp\ data from Fig.~\ref{quadv2} (left, solid triangles) are superimposed with their cubic trend (dashed). The solid curve is Eq.~(\ref{v22b}) rescaled to best match \pp\ and \auau\ data at lower \nch. The dash-dotted line $\propto \bar \rho_0^2$ is discussed in the text.
		Right:  Items in left panel transformed to $v_2 = \sqrt{V_2^2} / \bar \rho_0$ for comparison with Fig.~\ref{quadv2} (right).
	}  
\end{figure}

Figure~\ref{quadxx} (left) shows 62 and 200 GeV \auau\ quadrupole data in the form $V_2^2\{\rm 2D\} \equiv \bar \rho_0 A_Q\{\rm 2D\}$ vs charge density $\bar \rho_0$.  Those data correspond exactly to the open symbols in Fig.~\ref{alice4} (left). Also shown are 2.76 TeV \pbpb\ $V_2^2\ \equiv \bar \rho_0^2 v_2^2$ data (solid squares) based on $v_2\{4\}$  from Ref.~\cite{alicev2b} in Fig.~\ref{alice4} (left) and $\bar \rho_0$ data from Ref.~\cite{alicerho0} and 5 TeV \pbpb\ data (open squares) from Ref.~\cite{cmsabv2}.
\pp\ data from Fig.~\ref{quadv2}  (left), along with their dashed cubic reference trend, are superposed for comparison. 
Note that quadrupole correlation amplitude $V_2^2(\bar \rho_0)$ (as correlated pairs) increases by {\em almost six orders of magnitude} consistently across three collision systems. There is no rescaling of data to achieve that correspondence.

{Given the data trend in Fig.~\ref{quadxx} (left) the conjecture presented in Eq.~(\ref{ppconject}) may be tested. By analogy with the \pp\ result the quadrupole trend for \aa\ data with \aa\ geometry parameters is approximated by} (see Eq.~(\ref{rho0}))
\bea \label{v22b}
 V_2^2(n_{ch}) &\equiv& \bar \rho_0^2 v_2^2(n_{ch}) \propto \bar \rho_s(n_{ch}) \times \bar \rho_h(n_{ch})
\\ \nonumber
&\propto& (N_{part}/2) \bar \rho_{sNN} \times N_{bin}  \bar \rho_{hNN},
\eea
where geometry parameters are derived from a TCM analysis of 2.76 TeV \pbpb\  data reported in Ref.~\cite{tompbpb} and summarized in App.~\ref{spspec}. Equation~(\ref{v22b}), rescaled to best match \pp\ data at lower \nch, is represented by the solid curve in the left panel. The details are interesting.

Up to a transition point at $\bar \rho_0 \approx 15$,  first noted in Ref.~\cite{ppbpid}, \aa\ $N_{part}/2 \approx N_{bin} \approx 1$ due to {\em exclusivity}~\cite{tomexclude}. For resulting isolated \nn\ collisions one may then assume that $\bar \rho_{hNN} \propto \bar \rho_{sNN}^2$ as for \pp\ collisions~\cite{ppprd}. As a result, the \aa\ solid curve follows the dashed \pp\ cubic reference $\propto x^3$ {\em and describes the \auau\ data within that interval}. 

Above the transition point $\bar \rho_{xNN}$ trends are approximately constant for \aa\ collisions (see Fig.~\ref{pbpbglaub}, left) and the overall trend of  Eq.~(\ref{v22b}) is then dominated by product $N_{part} \times N_{bin}$. From Fig.~\ref{pbpbglaub} (right) $N_{part}\propto \bar \rho_0^{0.92}$ and for 200 GeV \auau\ collisions $N_{bin} \propto N_{part}^{4/3}$~\cite{powerlaw}. That combination leads to $N_{part} \times N_{bin} \propto \bar \rho_0^{2.15}$ approximated by the dash-dotted reference line $\propto x^2$ in the left panel. 

{To summarize, \aa\ $v_2(n_{ch})$ data in the form $V_2^2(n_{ch})$ follow a cubic power law (dashed line) on \nch\ consistent with \pp\ data below a transition point near $\bar \rho_0 = 15$ and a quadratic power law (dash-dotted line) above the transition point consistent with evolution of \aa\ geometry parameters. That confirms the relation $ V_2^2(n_{ch}) \propto \bar \rho_s \times \bar \rho_h$ as common to three A-B collision systems.}

{Only the accuracy {\em and consistency} obtained from model fits to 2D angular correlations~\cite{ppquad,anomalous,v2ptb} provide effective coverage over a sufficient range of \aa\ $\bar \rho_0$ to establish a smooth transition from \pp\ (and \aa!) cubic to \aa\ quadratic trend, confirming a transition at $\bar \rho_0 \approx 15$ in Eq.~(\ref{v22b}).}
That result is remarkable: Existence and implications of exclusivity in \pa\ and \aa\ collisions were  recently discovered~\cite{tomexclude,tomglauber,tompbpb}, whereas  \pp\ and \auau\ $v_2\{\text{2D}\}$ data used here were obtained ten to fifteen years ago. Yet the $v_2$ data effectively corroborate the exclusivity result. {   Exclusivity is discussed in App.~\ref{spspec} in relation to Fig.~\ref{pbpbglaub}. As a result of exclusivity \aa\ collisions are restricted to single \nn\ collisions up to a transition point near $\bar \rho_0 = 15$. The transition from dashed to dash-dotted trends in Fig.~\ref{quadxx} (left) is the result.}

{A claim could be made that power-law trends associated with $v_2 \rightarrow V_2^2$ (pairs) data as in Fig.~\ref{quadxx} (left) may arise from collective effects associated with a dense flowing medium. Since ``collective'' literally means ``correlated'' there is a variety of phenomena within collisions that may be called ``collective'' such as dijet production (correlation hard component), Bose-Einstein correlations, electron pairs from gamma conversions and unlike-sign pair correlations from projectile nucleon dissociation along the beam axis (correlation soft component)~\cite{anomalous}. For \pp\ collisions the soft component varies as $\bar \rho_s \approx \bar \rho_0$ and dijet production varies as $\bar \rho_s^2 \approx \bar \rho_0^2$ while the quadrupole component varies as $\bar \rho_s^3 \approx \bar \rho_0^3$ per Fig.~\ref{quadv2}. No compelling argument emerges to associate power-law trends as in Fig.~\ref{quadxx} (left) with a flow phenomenon.}

{Figure~\ref{quadxx} (right) shows $v_2(\bar \rho_0)$ data for 200 GeV \pp\ collisions, 62 and 200 GeV \auau\ collisions (all three via model fits to 2D angular correlations), 2.76 GeV \pbpb\ collisions via the $v_2$\{4\} method (solid squares) and 5 TeV \pbpb\ data from Ref.~\cite{cmsabv2} (open squares). The curves at right are  curves at left suitably transformed. The correspondence among \pp\ \{2D\} and \auau\ \{2D\} data within point-point uncertainties is notable. Also note the simplicity {\em and interpretability} of $V_2^2$ trends on $\bar \rho_0$ at left vs the complexity of the $v_2$ trend on the same variable. 

{For nominal flow measure $v_2$, maximum values for \aa\ are comparable to maximum values for \pp.} A ratio of two experimental results (quadrupole emission vs nucleon and parton fragmentation) hides essential information carried by angular correlations.}
{It is  interesting that 5 TeV \pbpb\ $v_2(p_t)$ data (open squares) from Ref.~\cite{cmsabv2} follow the 200 GeV \pp\ trend (dashed curve) with the 200 GeV \auau\ data (open circles) up to $\bar \rho_0 \approx 30$ confirming that both \aa\ systems are restricted within that interval to single \nn\ collisions by exclusivity.}

\subsection{Implications for large range of $\bf V_2^2$ values} \label{implications}

{Aside from establishment of a common quadrupole trend for \pp\ and \aa\ collisions in Fig.~\ref{quadxx} (left) it is also notable that the quadrupole correlation amplitude {\em increases by factor $10^6$} from lowest-\nch\ \pp\ to midcentral \pbpb. It is unlikely that such a large range could be manifested by changing properties (density, temperature, volume) of a dense flowing particle source within the context of Eq.~\ref{v2av}. It is more likely that the trend arises from a fundamentally different process.

Given the \pp\ cubic trend in Fig.~\ref{quadv2} (left) one may conjecture that quadrupole production arises from {\em individual} elementary three-gluon interactions. The quadrupole {\em production frequency} may then depend on the event-wise density of low-$x$ gluons, with density cubed determining the \nn\ quadrupole amplitude. Via a combination of fluctuating depth of splitting cascades in individual \nn\ collisions and \aa\ centrality variation the million-fold increase of mean correlation amplitude from \pp\ to central \aa\ may then be explained. {In that scenario there is no common reaction plane (as assumed for Eq.~(\ref{v2av})). Each quadrupole emission occurs independently.}

For low-multiplicity \nn\ collisions quadrupole hadron single emissions may occur within a small fraction of collisions. However, for higher-\nch\ \nn\ collisions, and especially more-central \aa\ collisions, multiple quadrupole interactions per event may occur. A similar scenario describes dijet (color dipole) production. For example, in Ref.~\cite{ppprd} the diject production probability per event within a limited $\eta$ acceptance for 200 GeV NSD \pp\ collisions is about 0.03 (~\cite{jetspec2} Sec.\ VI B), but because $\bar \rho_h \propto \bar \rho_s^2$ for \pp, and with an increase of event \nch\ by factor 10, jet production increases by a factor 100 leading to {\em multiple dijets per \pp\ event for larger \nch.} 
One should note that a sufficient number of quadrupole emissions {\em per event}, { with {\em independent azimuth orientations as for dijets}}, should result in reduction of a {\em net} quadrupole amplitude by averaging over azimuth, thereby accounting for decreases in Fig.~\ref{quadxx} (left) above $\bar \rho_0 \approx 200$.

\subsection{$\bf p_t$-integral quadrupole energy trends} \label{quadedep}

Figure~\ref{glaubtrend} (left) summarizes measured quadrupole energy dependence from AGS to LHC energies. $A_Q$ values reported previously are here rescaled per the description at the beginning of Sec.~\ref{modelfits}. 
$A_Q$ data maxima near $b/b_0 \approx 0.5$ minimize the relative effects of jet (``nonflow'') contributions to $A_Q\{\rm method\}$. 
Above 13 GeV the function $A_Q \propto \log(\sqrt{s_{NN}}/13~ \text{GeV})$ (solid line) describes the energy evolution with zero intercept at $13\pm 2$ GeV. 
Energy dependence below 13 GeV varies as $A_Q \propto \ln(\sqrt{s_{NN} }/ \text{3 GeV})$, corresponding to the Bevelac-AGS transition from ``squeeze-out'' (negative) to ``in-plane expansion'' (positive) $v_2$ values~\cite{squeezeout}.
In terms of per-particle measure $A_Q$ one observes a transition in the energy trend near 13 GeV, suggesting {\em two distinct physical mechanisms} for the azimuth quadrupole: plastic nucleon flow at lower energies and emergence of a new QCD three-body process among gluons at higher energies.

\begin{figure}[h]
     \includegraphics[width=1.65in]{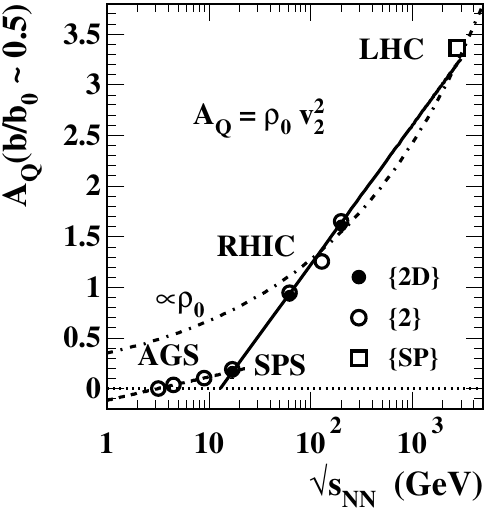}
    \includegraphics[width=1.65in]{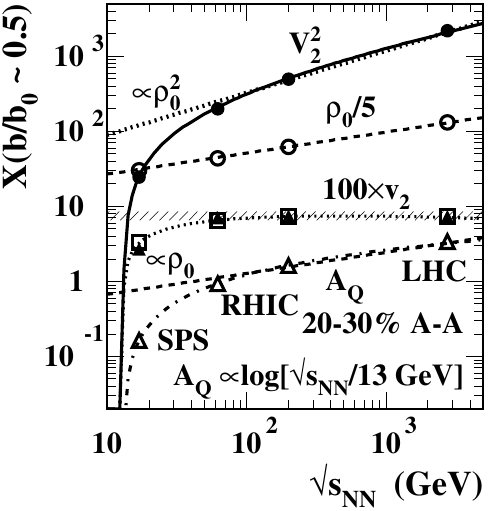}
\caption{\label{glaubtrend}
Left: Quadrupole amplitude $A_Q(n_{ch},\sqrt{s_{NN}})$ vs collision energy $\sqrt{s_{NN}}$ spanning three orders of magnitude from AGS to LHC. Two distinct trends logarithmic on collision energy emerge.
Right: Comparison of trends for $A_Q(n_{ch},\sqrt{s_{NN}})$ (open triangles, dash-dotted), $V_2^2(n_{ch},\sqrt{s_{NN}})$ (solid dots, solid), $v_2(n_{ch},\sqrt{s_{NN}})$ (solid triangles, dotted) and $\bar \rho_0$ (circles, dashed).
}  
\end{figure}

Figure~\ref{glaubtrend} (right) shows A$_\text{Q}$ values (open triangles) from the left panel in a log-log plot format. Also shown are charge density $\bar \rho_0/5$ at midrapidity (open circles) and quantity $V_2^2 = \bar \rho_0$A$_\text{Q}$ (solid dots). Measured $v_2$ multiplied by 100 (solid triangles) is also shown~\cite{v2ptb,alicev2b}.%
\footnote{Two RHIC $v_2$ values reported in Ref.~\cite{v2ptb} have been corrected by factor $\sqrt{2}$ as noted at the beginning of Sec.~\ref{modelfits}.}
The curves represent simple models for the measured quantities. The dash-dotted curve repeats the trend \mbox{A$_\text{Q}$ $\propto \log(\sqrt{s_{NN}}/13~ \text{GeV})$} from the left panel. The dashed line  represents  empirical trend $\bar \rho_0 \propto (\sqrt{s_{NN}})^{0.28} \rightarrow \exp[0.28 \log(\sqrt{s_{NN}})]$, an exponential increase with $\log(\sqrt{s_{NN}})$. The solid curve is $V_2^2 = \bar \rho_0$A$_\text{Q}$. The dotted curve is $v_2 = \sqrt{\text{A$_\text{Q}$}/\bar \rho_0}$, and the open squares represent $v_2$ predictions based on that relation which may be compared with $v_2$ data (solid triangles). Empirically, $A_Q(\sqrt{s_{NN}})$ (dash-dotted) approximates $\bar \rho_0$ variation (lower dashed) above 50 GeV {  (compare the $A_\text{Q}$ dash-dotted curve with the $\propto \bar \rho_0$ lower dashed curve, and see the equivalent comparison at left)}. As a result, $v_2(\sqrt{s_{NN}})$ (dotted curve and solid triangles) effectively saturates (hatched band) above that point. The quadrupole amplitude as measured by $V_2^2$ (solid, correlated pairs) continues to increase $\propto \bar \rho_0^2$ to highest energies. That trend is consistent with the \aa\ result in Fig.~\ref{quadxx} (left) above the transition at $\bar \rho_0 \approx 15$ where  $V_2^2 \propto \bar \rho_0^2$. Energy and multiplicity trends for $V_2^2$ are consistent but that important result is masked by ratio $v_2$.

{To summarize this section: 

Sec.~\ref{modelfits} introduces model fits to 2D angular correlations for 200 GeV \pp\ collisions that provide accurate determination of jet and quadrupole amplitudes measured by number of correlated pairs. Quadrupole measure $V_2^2$ is thus an {\em extensive} quadrupole measure plotted vs  charge multiplicity density $\bar \rho_s$ (soft) or $\bar \rho_0$ (total charge density). 

Section~\ref{alternative} considers a group of alternative $v_2$ measures, most developed in the context of Eq.~(\ref{v2av}) and assuming estimation of an event-wise reaction plane applied to an entire \aa\ collision. A contrast is made with model fits to 2D angular correlations that do not required a reaction-plane estimation.

Section~\ref{ppquadjet} describes measurements of jet and quadrupole amplitudes for 200 GeV \pp\ collisions and notes two trends -- dipoles (dijets) $\propto \bar \rho_s^2$ and quadrupoles $\propto \bar \rho_s^3$ -- that suggest color quadrupoles arise from three-body (three-gluon) interactions by analogy with color dipoles (dijets) generated by two-gluon interactions.

Section~\ref{multcent} demonstrates that quadrupole $V_2^2(\bar \rho_0)$ trends for 200 GeV \pp\ collisions, 200 GeV \auau\ collisions, 2.76 TeV \pbpb\ collisions and 5 TeV \pbpb\ collisions follow a common trend described by Eq.~(\ref{v22b}) over six orders of magnitude. 

Section~\ref{implications} notes certain implications from Fig.~\ref{quadxx} (left): Quadrupole production consists of individual three-gluon interactions. There would then be no reaction plane common to an entire \aa\ (or \pp) collision. Each quadrupole emission would  have an independent azimuth orientation as for dijets. Over a large range, quadrupole emission frequency may vary from single emission in a small fraction of events (low-\nch\ \pp) to multiple emissions per event (more-central \aa).

Section~\ref{quadedep} presents several quadrupole-related measures vs collision energy $\sqrt{s_{NN}}$. Conventional measure $v_2$ is observed to saturate at a fixed value for $\sqrt{s_{NN}} > 50$ GeV, whereas amplitude $V_2^2$ (correlated-pair measure) continues to increase rapidly over available energies. A primary reason for the discrepancy is the nature of $v_2$ as a {\em ratio} of two strongly-increasing quantities whose energy trends are distinct but similar and so nearly cancel.}

Given the simple $V_2^2$ (correlated-pair) trends on particle density $\bar \rho_0$ and collision energy $\sqrt{s_{NN}}$ there is no evidence for a varying (or any) thermodynamic equation of state or QCD phase transition from measured \v2\ data trends that remain simple and consistent from low-multiplicity \pp\ collisions to central \aa\ collisions (Fig.~\ref{quadxx}) and over a large energy interval (Fig.~\ref{glaubtrend}).}

\section{Systematic uncertainties} \label{syserr}

Uncertainties relating to previously analyzed 200 GeV \auau\ $v_2$ data, variation of $v_2(p_t,n_{ch})$ data with collision energy and variation of quadrupole spectrum structure and source boosts with collision energy are considered.

\subsection{200 GeV quadrupole spectra}

Systematic uncertainties for the analysis in Ref.~\cite{quadspec} were presented in that article. However, two further comments are appropriate:    Lambda data from 0-10\% central 200 GeV \auau\ collisions in Fig.~\ref{x2} (left) (solid points) were released after the analysis in Ref.~\cite{quadspec} was published. The significant negative values below the zero intercept near $y_t = 0.6$ confirm the {\em prediction} of the quadrupole-spectrum analysis represented by the dash-dotted curve. 

{   It is instructive to compare the statistical uncertainties and data trends in Fig.~\ref{x1} (left) of this article with those in Fig.~\ref{xbig}. 
In the latter case relative errors (on a semilog plot) are comparable for all \pt\ values except the last few points, and data values fall approximately exponentially. In the former case data values increase dramatically and errors for larger \pt\ are comparable to the panel height (for original data in that study) whereas the errors at smaller \pt\ are not visible. The difference is a consequence of the structure of Eq.~(\ref{combfac}) which may be expressed in the form
\bea
v_2(p_t) &\propto& [p_t' / \bar \rho_0(p_t)] \bar \rho_2(p_t)
\eea
in the lab frame. Referring to the 200 GeV pion spectrum in Fig.~\ref{alice5} (a) and the relevant \pt\ interval, the quantity in square brackets varies by five orders of magnitude. {\em Essential quadrupole boost information} carried by $v_2(p_t)$ data at lower \pt\ is thereby suppressed by that factor. A similar argument applies to $V_2^2(\bar \rho_0)$ vs $v_2(\bar \rho_0)$ data in Figs.~\ref{quadv2} and \ref{quadxx}. Primary information (power-law trends over six orders of magnitude) is carried by extensive measure $V_2^2(\bar \rho_0)$ representing total correlated pairs. As the square root of a pair ratio, $v_2(\bar \rho_0)$ tends to conceal that information while data uncertainties remain large.}

\subsection{$\bf v_2$ data: 200 GeV vs 2.76 TeV }

Figure~\ref{alice1} shows $v_2\{\rm SP\}(p_t,n_{ch})$ data for 15 million 2.76 TeV \pbpb\ events with statistical and systematic errors combined in quadrature. The error bars are much reduced from the 200 GeV data in Fig.~\ref{x1} (left) (e.g.\ 200 GeV kaon and Lambda data were based on $\approx 1.5$ million minimum-bias \auau\ collisions). However, the trend of errors is the same: $v_2$ errors at lower \pt\ are invisible suggesting that important information in that interval is {\em visually suppressed}, whereas \v2\ data transformed to quadrupole spectra make  information visually accessible.
Note that new 200 GeV pion data (solid triangles) in Fig.~\ref{alice2} (a), and especially Fig.~\ref{alice3} (a), and the dashed curves from Fig.~\ref{x1} (right) agree with \pbpb\ data within errors.

\subsection{2.76 TeV quadrupole spectra and source boosts}

{  Plotted error bars for the quadrupole-spectrum data in Fig.~\ref{alice8c} are simply published $v_2\{\rm SP\}(p_t,n_{ch})$ data uncertainties in Fig.~\ref{alice1} transformed just as for the data values. {Resulting error bars are typically smaller than the points.} For example,  pion error bars plotted  in Fig.~\ref{alice8c} have been increased by factor 3 but are still not visible.  For pion data there appears to be excellent systematic control, especially in relation to the 200 GeV quadrupole spectrum data (inverted solid triangles).
However, as noted elsewhere there are substantial low-\pt\ systematic deviations for kaon and proton data.}

The \nch\ and energy dependence of unit-integral $\hat S_2(m_t',T_2,n_2)$  defined by Eq.~(\ref{stuff2}) depends on parameters $T_2$ and $n_2$ determined at 2.76 TeV by spectra in Fig.~\ref{alice8c}. Presently-available data do not require any significant change in $T_2 \approx 93\pm 1$ MeV with either centrality or energy. Model exponent $n_2$ decreases significantly with energy (harder spectrum) from $n_2 = 14\pm 1$ at 200 GeV to $n_2 = 12\pm 1$ at 2.76 TeV evident in that figure.

That quadrupole source boost $\Delta y_{t0}(n_{ch})$ varies significantly with \aa\ centrality at 2.76 TeV is demonstrated by comparison of Figs.~\ref{alice3shift} and \ref{alice3}. An inferred centrality variation is sketched as the linear trend in Fig.~\ref{alice4} (right). A 20\% change in the slope of Fig.~\ref{alice4} (right) cannot be excluded by data, and the  trend could be significantly nonlinear on fractional cross section $\sigma/\sigma_0$. The \v2\ data are consistent with no significant energy dependence of $\Delta y_{t0}$ between 200 GeV and 2.76 TeV at the current level of uncertainty in inferred boost values. Presently-available \v2\ data do not require significant dispersion in the source boost for a given collision system (no evidence from \v2\ data for Hubble-like expansion of a bulk medium).

\section{discussion} \label{disc1}

This section considers  interdependence of three hadron production mechanisms contributing to $v_2(p_t,\bar \rho_0)$ as a ratio and the process of predicting $v_2(p_t)$ from a generic hydro model. It compares hydro predictions to $v_2(p_t,\bar \rho_0)$ and $V_2^2(p_t,\bar \rho_0)$ data, reviews details of a specific viscous-hydro theory in relation to $v_2(p_t)$ data and considers experimental evidence for monopole boost variation on \yt.

\subsection{Interrelation of production mechanisms}

According to some conventional \aa\ collision narratives~\cite{pbm,uliflow} almost all particles produced in high energy  collisions emerge from ``freezeout'' of a common flowing QCD bulk medium. That description may be contrasted with observed properties of three components distinguishable among final-state hadrons: (a)  soft (projectile-nucleon dissociation), (b) hard (scattered-parton fragmentation to jets) and (c) azimuth quadrupole radiation.

(a) The soft component of particle production for yields, spectra and correlations is accurately isolated per TCM analysis methods~\cite{ppprd,ppbpid}. Soft and hard yields are linked by a quadratic relation as one manifestation of {\em exclusivity} {  (see App.~\ref{spspec})}~\cite{ppprd,ppbpid,tomexclude}. The soft spectrum component is a Boltzmann exponential on transverse mass \mt\ with power-law tail. Soft-component properties are independent of collision \nch\ for \pp\ collisions or geometry for \aa\ collisions, consistent with soft hadron formation {\em outside a collision space-time volume}. Boltzmann slope parameter $T_0$ is $\approx 145$ MeV for pions and $\approx 200$ MeV for heavier hadrons. 
 Power-law exponent $n$ varies with collision energy per Gribov diffusion  within a parton splitting cascade~\cite{gribov} and is $\approx 8.8$ for 2.76 TeV.

(b) The hard component of particle production manifests as jets in yields, spectra and angular correlations. As noted, hard yields vary quadratically with soft yields for \pp\ collisions~\cite{ppprd} and \nn\ collisions within \aa\ collisions~\cite{tompbpb}. Spectrum hard components are consistent with jet formation per QCD collinear factorization~\cite{jetspec2}. Angular correlation structure manifests as a same-side (on azimuth $\phi$) 2D peak representing {\em intra}jet correlations and an away-side $\cos(\phi-\pi)$ 1D peak representing jet-jet correlations~\cite{anomalous,ppquad}. 
Unlike the soft component the hard-component spectrum shape may exhibit significant \nch\ or centrality dependence~\cite{pidpart2}. But modifications to spectrum hard-component structure conventionally attributed to ``jet quenching'' in a dense medium have been identified as possible consequences of \nn\ exclusivity and parton relativistic time dilation~\cite{tompbpb}.

(c) Some detailed properties of the quadrupole component are newly reported in the present study. Like the spectrum soft component, the quadrupole spectrum is a Boltzmann exponential on \mt\ with power-law tail, but with slope parameter $T_2\approx 95$ MeV common to several hadron species and power-law exponent $n \approx 12$ for 2.76 TeV collision energy. These values are markedly different from those for the SP spectrum soft component.

The three components may be described as {\em distinct but interrelated}, which should not be surprising given that jet production and soft-particle production in \pp\ collisions are already precisely related by the empirical quadratic relation $\bar \rho_h \propto \bar \rho_s^2$. In the present study those interrelations are further elaborated, especially the results demonstrated by Fig.~\ref{quadxx} (left) and the discussion in Sec.~\ref{multcent}. Given new quadrupole details it is quite unlikely that almost all hadrons emerge from a common dense medium, or that a hydrodynamic description is relevant.

\subsection{Predicting $\bf v_2(p_t)$ data from a hydro model} \label{hydropredict}

Given Eq.~(\ref{combfac}) the general structure of $v_2(p_t)$ follows as
\bea \label{v2gen}
v_2(p_t,\Delta y_{t0},\Delta y_{t2}) &\approx& p'_t(p_t,\Delta y_{t0}) \frac{ \Delta y_{t2}}{2T_2} \frac{\bar \rho_2(p_t,\Delta y_{t0})}{\bar \rho_0(p_t)},~~~~
\eea
where $\Delta y_{t2}$ represents a quadrupole boost.
In effect, three particle sources are included in that relation: (a) the quadrupole spectrum $\bar \rho_2(p_t)$, established via the present analysis and Ref.~\cite{quadspec} as representing a unique particle source, and $\bar \rho_0(p_t)$ representing (b) soft (projectile nucleon dissociation) and (c) hard (parton fragmentation to jets) particle production as reported in Refs.~\cite{ppprd,fragevo,ppbpid} among others. 
The general shape of $v_2(p_t)$ vs \pt\ is simply explained by that combination. Factor $p_t'$ causes strong linear rise through a zero intercept at lower \pt, but plotting on linear \pt\ obscures a shift of $\bar \rho_2(p_t)$ in the lab by $\Delta y_{t0}$ (see Fig.~\ref{alice1xx} (a)). 
The dominant {\em jet contribution} to $\bar \rho_0(p_t)$ at higher \pt\ causes a strong reduction of $v_2(p_t)$ above $p_t \approx 3$ GeV/c as discussed below.

If the NJ quadrupole spectrum $\bar \rho_2$ were equivalent to  SP spectrum $\bar \rho_0$ as assumed for Eq.~(\ref{v2av}), {\em possibly} justifying the claim $v_2 \sim \langle \cos(2 \phi_r) \rangle$, Eq.~(\ref{combfac}) then simplifies to
\bea \label{v2simple}
v_2(p_t,\Delta y_{t0},\Delta y_{t2}) &\approx& p'_t(p_t,\Delta y_{t0}) \frac{\Delta y_{t2}}{2T_2} .
\eea
That ``ideal hydro'' trend is shown below in a conventional $v_2(p_t)$ vs \pt\ plot format and in a modified  format.

Figure~\ref{schema1} (left) shows Eq.~(\ref{v2simple}) for three hadron species ($\pi$, K, p) and fixed $\Delta y_{t0} = 0.6$ (based on Ref.~\cite{quadspec}). Expression ${\Delta y_{t2}}/{2T_2} \approx 0.15$/GeV is adjusted so that the ``ideal hydro'' trends (solid, dashed, dash-dotted) correspond approximately to $v_2(p_t)$ data at lower \pt\ in Fig.~\ref{x1} (left) (actual values are 0.17, 0.15 and 0.14 for pions, kaons and protons). $v_2(p_t)$ data suggest that $T_2 \approx 95$ MeV; ratio $\Delta y_{t2}/\Delta y_{t0} \approx 0.05$ implies  $f(y_t,\Delta y_{t0},\Delta y_{t2})$ from Eq.~(\ref{stuff}) deviates from unity by only a few percent over a relevant \yt\ interval and may be ignored. 

{The bold dotted curve is a viscous hydro prediction for protons from 2.76 TeV \pbpb\ collisions~\cite{shen}. Note that whereas the ``ideal hydro'' curves descend linearly to negative values at lower \pt\ the hydro theory $v_2(p_t)$ trend remains at or above zero {} down to zero \pt. {} The ``viscous'' aspect of viscous hydro (deviation from ``ideal'') appears to emerge in  this case only above 2.5 GeV/c.

\begin{figure}[h]
 \includegraphics[width=1.65in,height=1.65in]{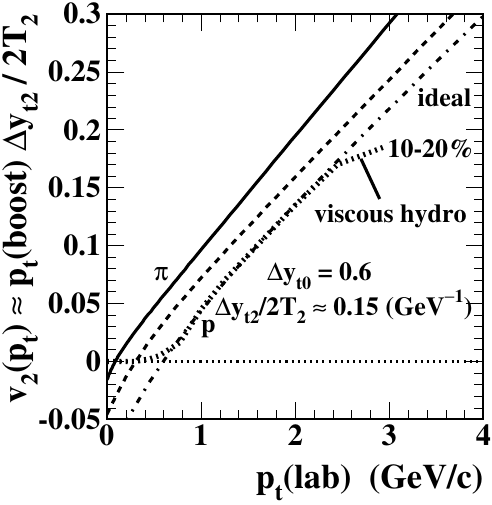}
  \includegraphics[width=1.65in,height=1.625in]{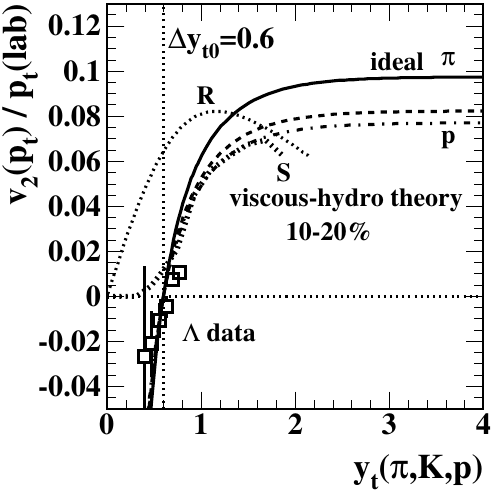}
\caption{\label{schema1}
Left: $v_2(p_t)$ trends for three hadron species vs $p_t(\text{lab})$ (solid, dashed, dash-dotted) assuming ``ideal hydro'' conditions and a monolithic flowing-bulk-medium hadron source. {   The bold dotted curve is a viscous-hydro theory prediction for protons from  central 2.76 TeV \pbpb~\cite{shen}.}
Right: Ratio $v_2(p_t) / p_t(\text{lab})$ vs transverse rapidity $y_t$ defined for each hadron species.  ``Ideal-hydro'' curves (solid, dashed, dash-dotted) have a common form given by Eq.~(\ref{kine}). {   Viscous-hydro predictions are shown for a previous prediction (light dotted, R)~\cite{rom} and a more-recent prediction (bold dotted, S)~\cite{shen}.} The open squares are Lambda data for 0-10\% central \auau\ collisions~\cite{newstarpion}.
 } 
 \end{figure}

Figure~\ref{schema1} (right) shows ratio $v_2(p_t) / p_t$(lab) vs proper transverse rapidity for each hadron species. ``Ideal'' curves have a universal form that increases through zero at $\Delta y_{t0}$ and corresponds to Fig.~\ref{boost3} (right) of App.~\ref{boost}. {  Also shown are an earlier viscous-hydro theory prediction for protons from \auau\ collisions (light dotted R)~\cite{rom} and a more-recent prediction for \pbpb\ (bold dotted S)~\cite{shen}.}
{Beyond \yt\ = 1.5 the viscous-hydro model curve {\em for this ratio measure} descends}, possibly to emulate $v_2(p_t)$ data trends {  (note that within that interval the theory $v_2(p_t)$ trend itself is monotonically increasing as in the left panel)}. 
{\em By construction} $v_2$ includes single-particle spectrum $\bar \rho_0$ in its denominator. {That factor can be removed} as in Sec.~\ref{200gevquad} to determine what remains. The result is a universal quadrupole spectrum in the boost frame consistent with a Boltzmann exponential on transverse mass with quite low slope parameter $T_2 \approx 95$ MeV. 

That $v_2(p_t)$ data drop sharply away from the ideal-hydro trends toward zero has been attributed to viscosity of a bulk medium assuming that almost all hadrons emerge from that common medium~\cite{songperliq}. But the fall-off may be more simply explained by quadrupole spectra quite different from  SP spectra (that describe most hadrons) in  Eq.~(\ref{v2struct}). 
Note that in Fig.~\ref{xbig} the power-law tail for 200 GeV quadrupole function $\hat S_2(m_t)$ has exponent $n_2 = 14$, with $n_2 = 12$ for 2.76 TeV in Fig.~\ref{alice8c}. Also in Fig.~\ref{alice8c} the exponent for SP spectrum soft component $\hat S_0(m_t)$ is $n_0 = 8.8$. SP spectrum hard components, as densities on transverse rapidity \yt, are well described by a Gaussian with {\em exponential} tail. The latter is equivalent to a {\em power-law} tail on \pt\ with exponent $n_h \approx $ 6-7~\cite{tompbpb} (see its Table III and $n_h  \approx q + 2.2$). The smaller $n$ is the slower the falloff with increasing \pt.  In Eq.~(\ref{v2struct}) {\em ratio} the hard component in the denominator is responsible for the rapid falloff of $v_2(p_t)$ data. A significant jet contribution in the numerator (nonflow) might reduce the $v_2(p_t)$ falloff rate and be mistaken as arising from a medium viscosity.

\subsection{Viscous-hydro theory for Pb-Pb collisions} \label{shen}

Reference~\cite{shen} reports viscous-hydrodynamic predictions for \pt\ spectra and $v_2(p_t)$ from \pbpb\ collisions at the LHC related to RHIC results. It cites ``Compelling evidence for fluid dynamical  behavior of...collisions....'' referring in part to RHIC ``white papers'' published in 2005 and interpreted  to support claims of ``perfect fluid'' formation in \auau\ collisions~\cite{perfect}. Reference~\cite{shen} seeks to determine ``how good this agreement [between hydro predictions of v2 and data] is quantitatively....''

The assumed context closely coincides with that for Eq.~(\ref{v2av}): Almost all hadron production arises from a dense flowing medium, i.e.\ a {\em monolithic} collision model. A theoretical model of hadron production based on flows should then describe hadron \pt\ spectra and angular correlations (in the form of $v_2(p_t)$). It is further assumed that a hydro description may only describe ``soft'' hadron production below some upper limit $\approx 3$ GeV/c, i.e.\ not extending  ``beyond the \pt\ range where the hydrodynamic description is expected to begin to break down due to the increasing influence of hard production processes...~\cite{shen}.''

Identified-hadron \pt\ spectra are fitted {\em quantitatively} with a blast-wave (BW) model over a limited \pt\ interval {\em often determined by agreement with data}~\cite{noblast}. Hydro model fits to RHIC pion spectra are nevertheless poor as in Fig.~2 (b) of Ref.~\cite{shen}. 
Radial flow is inferred {\em qualitatively} via ``flatness'' of \pt\ spectra: ``The LHC spectra are visibly flatter than at RHIC energies, reflecting stronger radial flow.'' But jet fragment distributions peak near 1 GeV/c~\cite{jetspec2} and exhibit strong collision-energy dependence, with jet fragment density $\bar \rho_{hNN} \approx \alpha \bar \rho_{sNN}^2$ and with $\alpha(\sqrt{sNN})$ as in Ref.~\cite{alicetomspec} and $\bar \rho_{sNN} \propto \log(\sqrt{s_{NN} / \text {10 GeV}})$. Thus, spectrum  evolution described as ``flattening'' is  dominated by jet production.

Identified-hadron $v_2(p_t,n_{ch})$ data are also fitted over limited \pt\ and \nch\ intervals. As with SP spectra, jets contribute strongly to angular correlations. Analysis methods denoted by $v_2\{2\}(p_t)$ or $v_2\{\text{EP}\}(p_t)$ are maximally sensitive to jets. Others may be less so to an ill-defined extent. In contrast, $v_2(p_t)$ from 2D model fits as in Refs.~\cite{anomalous,v2ptb} show no significant sensitivity to jet contributions. In Ref.~\cite{shen} Fig.~5 hydro theory curves favor \{2\} and \{EP\} data with substantial jet contributions.

{A critical issue for hydro theory is the \pt\ range over which it describes $v_2(p_t)$ data closely and how that compares with the range for a simpler description as in Sec.~\ref{quadspecmeth}.}
In Ref.~\cite{shen} Fig.~7 panel (b) (pions) hydro theory curves for four centralities continue to increase up to 3 GeV/c whereas STAR pion data in Fig.~\ref{x1} (left) of this study maximize near 2 GeV/c and then decrease. 
The solid curve in Fig.~\ref{x1} (left), derived from a universal quadrupole spectrum curve in Fig.~\ref{xbig}, continues to follow pion data  closely {up to 6 GeV/c}. The {\em same} curve in Fig.~\ref{alice2} (a) (now dashed) follows LHC pion data also up to 6 GeV/c. 

{  In the present study a theory curve from Ref.~\cite{shen} Fig.~7 (c) for protons from 10-20\% central \pbpb\ collisions is compared with various results from RHIC and LHC in Figs.~\ref{x1} (right), \ref{x2} (left), \ref{alice3} (c) and \ref{schema1}. The proton theory curve generally agrees with data {from \yt\ = 0.6 (data zero intercept) up to \yt\ $\approx$ 1.9 (3 GeV/c).} 
However,  hydro theory for protons remains {\em at or above zero} from the origin up to $p_t \approx 0.55$ GeV/c that corresponds to $\Delta y_{t0} \approx 0.6$ in Fig.~\ref{x1} (right) of the present study. While the inflection point on \pt\ in Ref.~\cite{shen} agrees with the present study, {\em lack} of a negative trend below that point is {substantially} different from RHIC Lambda data trends as in Fig.~\ref{x2} (left) and LHC proton data trends as in Fig.~\ref{alice3} (c) or Fig.~\ref{booststudy} (right).
Also note that above \yt\ = 1.9 (3 GeV/c), the dashed model curve in Fig.~\ref{alice3} (c) based on Eq.~(\ref{alice3}) describes $v_2(p_t)$ data accurately up to 6 GeV/c (\yt\ = 2.72) and continues smoothly beyond the data acceptance upper limit as a prediction.}

The viscous hydrodynamic model of Ref.~\cite{shen} includes the following assumptions (and associated parameter values): (a) equation of state, (b) freeze-out temperature, (c) chemical composition at freeze-out, (d) starting time and (e) viscosity. 
{Referring to its Fig.~3 for nonPID charged hadrons ``...viscous hydrodynamics gives an excellent description of the...data, even up to 3 GeV...(i.e.\ beyond the \pt\ range where the hydrodynamic description is expected to begin to break down due to...hard production processes and large uncertainties in the viscous correction....).''}
The complex {and limited} model reported in Ref.~\cite{shen} may be contrasted with the simple model presented in Sec.~\ref{quadspecmeth} that describes all available $v_2(p_t)$ data well. 
Given accurate descriptions over entire \pt\ acceptances with a simple model there seems to be no need for a complex hydro model applied to  restricted \pt\ intervals.

{This subsection addresses specific hydro theory results from Ref.~\cite{shen} as a reference. Other theory results may differ markedly.} Results for viscous-hydro Monte Carlos may be contrasted with results from TCM-based SP spectra and quadrupole spectra as in the present study. The TCM has by now been applied to a broad range of collision systems and hadron species~\cite{ppprd,hardspec,fragevo,jetspec2,tomalicempt,alicespec,mbjets,ppbpid,pidpart1,pidpart2,noblast,pppid,ppbnmf,tcmcompare,tompbpb}. Spectrum data are described over complete \pt\ acceptance ranges with model components that are consistent across multiple systems. The TCM is not derived from fits to individual spectra.

\subsection{Monopole boost distribution measurement} \label{monboost}

Based on Eq.~(\ref{combfac}), if $\bar \rho_2(p_t) \propto \bar \rho_0(p_t)$ consistent with the assumptions underlying Eq.~(\ref{v2av}) then $v_2(p_t) \propto p_t' = p_t$(boost) as in Eq.~(\ref{v2simple}), described in figures above as ``ideal hydro''. $v_2(p_t)$ data are indeed consistent with that relation up to 2 GeV/c as in Fig.~\ref{x1}. When data are plotted vs transverse rapidity \yt, as in Fig.~\ref{x1} (right) or Fig.~\ref{x2} (left), the data are found to approach or pass through a single zero intercept that then defines a single value for monopole boost $\Delta y_{t0}$. As noted in Sec.~\ref{quadspecmeth}, $v_{2i}(p_t)$ data might conflict with that simple model to reveal a monopole source-boost {\em distribution} on $\Delta y_{t0}$ corresponding to Hubble-like expansion of a dense bulk medium. {  An example simulation for protons is given in Fig.~12 of Ref.~\cite{lisa}. That result is similar to  taking the proton curve (dash-dotted) in Fig. 16 (left) and convoluting it with a broad boost ($\Delta y_{t0}$) distribution.} 
However, such deviations are not observed for RHIC and LHC data considered in the present study. {  That exercise calls into question whether a broad velocity (boost) distribution may be supported by data.}

Figure~\ref{booststudy} shows data from Figs.~\ref{x2} (left) for \auau\ and 
Fig.~\ref{alice3shift} (right) for \pbpb\ to provide a more-detailed study of structure near the zero crossing. The proton data appearing at right are unshifted, as opposed to the data in Fig.~\ref{alice3} (c). The bold dash-dotted curve in each panel is the same curve appearing in Fig.~\ref{x1} (right) and Fig.~\ref{x2} (left) derived by back-transforming the Boltzmann exponential in Fig.~\ref{xbig} for \auau\ applied here as well to \pbpb\ data. The hatched band shows $\Delta y_{t0} \approx 0.6$ inferred from \auau\ data. The vertical dotted line indicates the lower bound of the \pt\ acceptance in each case. {For two most-central event classes proton data are plotted as points with errors (statistical and systematic combined in quadrature as in Fig.~\ref{alice1}) to indicate significance of negative values below the crossover.}

{The statistical significance of negative-going proton data may be questioned in comparison to theory curve S (bold dotted). Based on uncertainties from Ref.~\cite{alicev2ptb} as plotted in Fig.~\ref{alice1} the mean values of proton data from two most-central event classes in Fig.~\ref{booststudy} (right) are $-0.75 \pm 0.33$ for four ``nonzero'' entries or $-0.50 \pm 0.22$ for six entries within the interval from acceptance threshold to $y_t = \Delta y_{t0} = 0.6$. One may also include information from Lambda data in the left panel and data following {\em predicted trends} (dash-dotted curves) over a large \pt\ range to conclude that the probability of a null hypothesis (no negative low-\pt\ trend as observed for theory curve S in the right panel) is at most few percent.}

 \begin{figure}[h]
  \includegraphics[width=1.65in,height=1.65in]{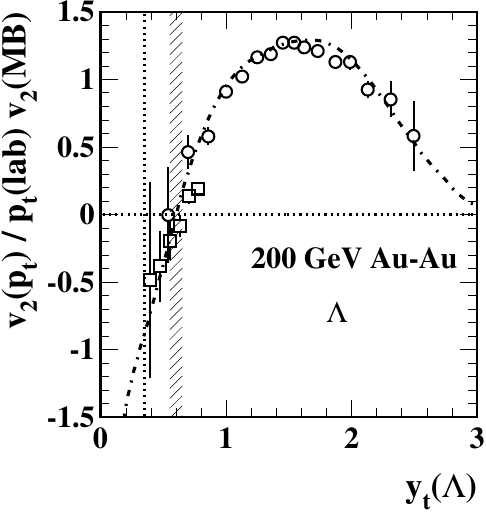}
   \includegraphics[width=1.65in,height=1.65in]{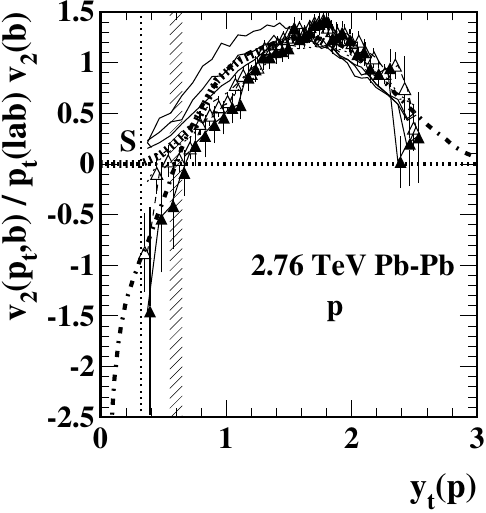}
\caption{\label{booststudy}
Left: Figure~\ref{x2} (left) repeated for comparison with $v_2(\text{ MB})=0.055$.
Right:  Fig.~\ref{alice3} (c) with proton data not shifted. {Data for more-central events { (class 1 solid, class 2 open)} appear consistent with factor $p_t' = p_t$(boost) as in Eq.~(\ref{combfac}) and thus do not {\em require} variation of monopole boost $\Delta y_{t0}$ within an event class.} {Event class 2 (5-10\% central) is  close to the dash-dotted trend, with smaller uncertainties, and is included as open crosses in Fig.~\ref{alice8ab} (left) for that reason.}
Those data demonstrate that estimation of { the width of a monopole boost distribution} relies on $v_2(p_t)$ data within a limited \pt\ interval. 
{Bold dotted curve S is a hydro theory prediction from Ref.~\cite{shen} extending up to 3 GeV/c (\yt\ = 1.9).}
 } 
 \end{figure}

Measurement of a boost distribution involves examining the structure of $v_2(p_t)$ data near {\em and below} inferred $\Delta y_{t0}$ on proper transverse rapidity for each hadron species. Employment of that specific plotting context is not conventional, and the relevant structure is usually referred to nonquantitatively as ``mass ordering.'' Thus, the only measurement of a boost-velocity variation appears in Ref.~\cite{quadspec} and the present study. In Fig.~\ref{booststudy} there is no detailed evidence for significant boost variation (per Hubble-like expansion) within a given event class. Although there is evidence of a significant variation of fixed $\Delta y_{t0}$ {\em across} event classes as in Fig.~\ref{alice4} (right) the shape of $v_2(y_t)$ remains consistent with factor $p_t' = p_t$(boost) over available data above a \pt\ acceptance boundary. {  The model described in Sec.~\ref{quadspecmeth}, with fixed $\Delta y_{t0}$, provides a comprehensive and accurate description of $v_2(p_t)$ for two collision systems over the full \pt\ range from zero up to 6.5 GeV/c. Data thus do not {\em require} any boost variation within a given event class. Limitations to such a determination (narrow \pt\ window) are apparent from Fig.~\ref{booststudy}.}

\section{Summary} \label{summ}

This article presents analysis of identified-hadron (PID) \pt-differential $v_2(p_t,n_{ch})$ data from 2.76 TeV \pbpb\ collisions to derive {\em quadrupole spectra}  associated with an azimuth quadrupole component of angular correlations. The procedure has been applied previously to 200 GeV \auau\ $v_2(p_t,n_{ch})$ data. The main goal of this analysis is to extract all available information from $v_2(p_t)$ data without {\em a priori} assumptions and to employ that information so as to better understand the mechanism that generates the quadrupole component and its properties. 

The $v_2$ measure itself presents multiple difficulties in its implementation and interpretation as described in this study. A preferred alternative is $V_2^2 \equiv \bar \rho_0^2 v_2^2$ (where $\bar \rho_0$ is a particle density near midrapidity), an {\em extensive} correlation measure proportional to number of correlated pairs.

Several novel findings have emerged as follows:

Detailed study of $v_2(p_t)$ algebraic structure in the context of a Cooper-Frye formalism reveals that $v_2(p_t)$ includes a factor $p_t'$ (in the source boost frame) that makes data interpretation ambiguous. Alternative format $v_2(y_t)/p_t$ vs transverse rapidity \yt\ (properly defined for each hadron species) reveals a {\em monopole boost} $\Delta y_{t0}$ that might correspond to radial flow in some context.

Transformation of  $v_2(p_t)$ data in a succession of steps from $ v_2(p_t)$ vs \pt\ to $\bar \rho_0(m_t') v_2(m_t')/p_t'$ vs $m_t'$ in a particle-source boost frame reveals that quadrupole spectra in that boost frame have the same shape for any hadron species. As a result, the properties of a particle source {\em unique to the quadrupole component} are inferred.

Analysis of 2.76 TeV \pbpb\ $v_2(p_t)$ data reveals that those $v_2(p_t)$ data trends do not depart significantly from 200 GeV trends. Source boost $\Delta y_{t0}$ exhibits a significant event-class \nch\ (centrality) dependence, but the 2.76 TeV mean value is consistent with the 200 GeV value.

A result of major importance emerges: quadrupole {\em amplitudes} in the form $V_2^2(n_{ch})$ obtained by model fits to 2D angular correlations (thus minimizing jet or ``nonflow'' contributions), previously obtained for 200 GeV \pp\ collisions and \auau\ collisions, are observed to follow a {\em common algebraic trend} first inferred for \pp\ collisions: a simple product of soft component (participant gluons or nucleons) times hard component (gluon or nucleon binary collisions) particle densities. The trend consists of two power laws: cubic for lower $n_{ch}$ (where gluons within single \nn\ collisions dominate) and quadratic for higher \nch\ (where multiple \nn\ collisions dominate). The transition from one trend to the other occurs at a value of charge density $\bar \rho_0$ previously related to {\em exclusivity} (a projectile nucleon may interact with only one target nucleon {\em at a time}). Thus, $v_2$ data trends correspond closely to an \nn\ constraint revealed by jet production systematics. The trend varies precisely over six orders of magnitude.

Issues for the $v_2$ measure in connection with quadrupole production include the following:
(a) $V_2^2$, as an {\em extensive} measure of quadrupole amplitude in terms of number of correlated pairs, increases strongly with collision energy whereas $v_2$, as a ratio of two strongly varying quantities, saturates above  50 GeV giving a misleading impression of the quadrupole production mechanism.
(b) $v_2(p_t)$ includes a ``hidden'' factor ($p_t$ in the source boost frame) leading to ill-defined inferences (e.g.\ ``mass ordering'') derived from $v_2(p_t)$ trends.
(c) Whereas $v_2(p_t)$ data may seem to be compatible with some hydrodynamic theory predictions because of its algebraic structure, a simple reformulation of $v_2(p_t)$ data related to extraction of quadrupole spectra reveals that hydro theories based on Hubble-like expansion of a bulk medium appear to be excluded.
(d) Various ``scaling'' strategies attempting to force $v_2(p_t)$ data for several hadron species onto a single locus are motivated by the ratio structure of that measure. The same data transformed to quadrupole spectra within a common boost frame reveal that the data trends then {\em follow nearly identical spectrum shapes}.

Taken together the results of this study, combined with related previous findings, appear to demonstrate that the azimuth quadrupole  source is a {\em distinct particle production mechanism} with unique characteristics that may be derived from $v_2(p_t,n_{ch})$ data in combination with other analysis. The nature of the source mechanism is suggested by data from 200 GeV \pp\ collisions: a QCD three-gluon interaction. A very similar data trend for \pbpb\ collisions suggests that the mechanism is universal within high-energy A-B collisions. Given overall data trends it is unlikely that the quadrupole component includes most hadrons emerging from Hubble-like expansion of a bulk medium as is conventionally assumed. It is more likely that the quadrupole component is ``carried'' by a small minority of final-state hadrons. Hydrodynamics-based theoretical descriptions appear unjustified by $v_2(p_t)$ data.

\begin{appendix}

\section{Boosted hadron sources} \label{boost}

This appendix reviews relativistic kinematics relating to nearly-thermal spectra for hadrons emitted from a moving (boosted) source based on the Cooper-Frye description of rapidly-expanding particle sources~\cite{cooperfrye}. Only azimuth-monopole and -quadrupole $p_t$ and $y_t$ spectrum components are considered. For simplicity ``thermal'' spectra are described in the boosted frame by Boltzmann exponentials on $m_{ti} - m_i$ (hadrons $i$). Relative hadron abundances are assumed to correspond to a statistical model, but not necessarily because of a thermalization process~\cite{statmodel}. The spectrum description on $m_t$ may be generalized to add a power-law tail for more accurate modeling of data~\cite{ppprd,hardspec,ppbpid}. The model provides a general description of hadron production from a source including (but not restricted to) a radially-boosted component with azimuth variation. This material is revised from Ref.~\cite{quadspec}.

\subsection{Radial boost kinematics} \label{boosteq}

The four-momentum components of a boosted source are first related to transverse rapidity $y_t \equiv \ln[(p_t + m_t)/m]$ for hadrons of mass $m$ and $m_t^2 = p_t^2 + m^2$. The boost distribution is assumed to be a single value $\Delta y_t$ for simplicity.  The {\em particle} four-momentum components are $m_{t} = m \cosh(y_{t})$ and $p_t = m\sinh(y_{t})$. The {\em source} four-velocity (boost) components are $\gamma_t = \cosh(\Delta y_t)$ and $\gamma_t \, \beta_t= \sinh(\Delta y_t)$, with $\beta_t= \tanh(\Delta y_t)$. Boost-frame variables for a hadron species with mass $m$ are defined in terms of lab-frame variables by
\bea \label{boostkine}
m'_t  &\equiv& m\, \cosh(y_t - \Delta y_{t}) =\gamma_{t}\, (m_t - \beta_{t}\, p_t)  \\ \nonumber
 &=& m_t\, \gamma_{t} \{ 1 - \tanh(y_t)\, \tanh(\Delta y_{t})  \} \\ \nonumber
p'_t &\equiv&   m\, \sinh(y_t - \Delta y_{t}) =\gamma_{t}\, (p_t - \beta_{t}\, m_t) \\ \nonumber
&=& m_t\, \gamma_{t} \{  \tanh(y_t) - \tanh(\Delta y_{t})  \}.
\eea
  
Fig.~\ref{boost3} (left) relates $p'_t \rightarrow p_t(\text{boost})$ to $p_t \rightarrow p_t(\text{lab})$. The main source of ``mass ordering'' for $v_2(p_t)$ at smaller $p_t$ (lower left), commonly interpreted to indicate ``hydro'' behavior, is a simple kinematic effect. The mass systematics hold for any boosted, approximately-thermal hadron source independent of boost mechanism (i.e.\ hydrodynamics is not required). {\em Zero intercepts} ($p'_t = 0$) of the three curves, denoted by $p_{t0} = m \sinh(\Delta y_{t})$, are relevant for discussion of the hydro interpretation of $v_2(p_t)$. 

{ A flowing (boosted) particle source might produce a ``mass ordering'' effect for some observables. $v_2(p_t)$ for different hadron masses might be shifted differently on \pt\ as in Fig.~\ref{x1} (left). But a data feature suggesting mass ordering does not guarantee hydrodynamic flow of a dense medium. In order to determine the nature of the source boost the behavior of various hadron species (masses) should be resolved on {\em proper transverse rapidity for each  species} as in Fig.~\ref{x1} (right). Some other QCD-related mechanism might be a more likely cause given evidence as in Figs.~\ref{quadv2} and \ref{quadxx}.}

 \begin{figure}[h]
   \includegraphics[width=1.65in,height=1.65in]{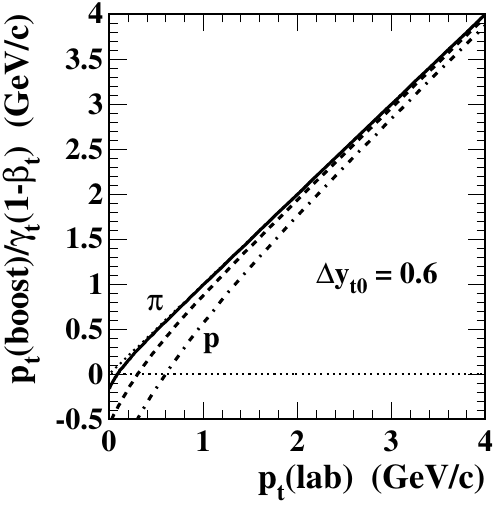}
   \includegraphics[width=1.65in,height=1.63in]{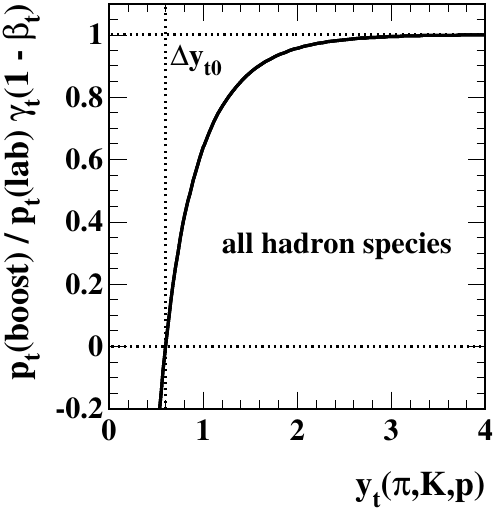} 
\caption{\label{boost3}
Left: $p'_t$ ($p_t$ in the boost frame) {\em vs} $p_t$ in the lab frame. The normalizing factor $\gamma_{t}(1 - \beta_{t})$ in the denominator insures that the combination approaches $p_t$ for large $p_t$. Note Lambda data in Fig.~\ref{x2} (left).
Right:  The quantity in the left panel divided by $p_t$(lab) vs proper $y_t$ for each hadron species demonstrating that Eq.~(\ref{kine}) is a universal trend common to all hadron species for given $\Delta y_{t0}$.
 } 
 \end{figure}

Figure~\ref{boost3} (right) relates ratio $p'_t / p_t$ to transverse rapidity $y_t(\pi,K,p)$ and illustrates one reason why plots on $y_t$ are a major improvement over $p_t$ or $m_t$. Rescaled $p'_t / p_t$\footnote{by factor $1/\gamma_{t}(1 - \beta_{t})$}  
\bea \label{kine}
 \frac{p'_t}{p_t\, \gamma_{t}(1 - \beta_{t})} 
    &=& \frac{1 - \beta_t /\tanh(y_t) }{1 - \beta_t}
\eea
increases from zero at monopole boost $\Delta y_{t0} $ and follows a {\em universal curve on $y_t$} to maximum 1 for {\em any hadron species}. Thus, rescaled $p'_t/p_t$
goes asymptotically to 1 for large $p_t$ (or $y_t$) independent of boost.
The simplified blast-wave (boost) model~\cite{cooperfrye}, invoked here for illustration, assumes longitudinal-boost-invariant normal emission from an expanding thin cylindrical shell~\cite{hydro2} with hadron mass $m$ and slope parameter $T$.  Boosted spectra on $y_t$ and $m_t$ are
\bea \label{boostx}
\rho(y_t,T,\Delta y_t)  \hspace{-.05in} &=&  \hspace{-.05in} A_{y_t}\,\exp\{ -m\, [ \cosh(y_t - \Delta y_t) - 1]/T\} \\ \nonumber
\rho(m_t,T,\beta_t) \hspace{-.05in} &=&  \hspace{-.05in} A_{m_t}\, \exp\{- [\gamma_t\, (m_t - \beta_t\, p_t) - m_i]/T \},~~~~~
\eea
providing a simplified description of ``thermal'' radiation from a radially-boosted cylindrical  source. Application of Eq.~(\ref{boostx}) requires a specific radial-boost model $\Delta y_t(r,\phi)$.

\subsection{Radial-boost models} \label{boostc}

In high-energy nuclear collisions there are at least two possibilities for the radial-boost model: (a) a monolithic, thermalized collectively-flowing hadron source (``bulk medium'') with complex transverse flow (source-boost) distribution on space $(r,\phi,t)$ dominated by monopole (radius-dependent Hubble-like expansion) and quadrupole (elliptic flow) azimuth components; or (b) several hadron sources, some with azimuth-modulated transverse boost. Hadrons may emerge from a radially-fixed source (soft component), from parton fragmentation (hard component), and possibly from a source{   (or sources)} with radial boost varying smoothly with azimuth including monopole and quadrupole components. 

An eventwise radial boost distribution with monopole and quadrupole components may be represented by
\bea \label{quadboost}
\Delta y_{t}(\phi_r) &=& \Delta y_{t0} + \Delta y_{t2}\, \cos(2 \phi_r ) \\ \nonumber
\beta_t(\phi_r) &=&  \tanh[\Delta y_t(\phi_r)] \\ \nonumber
&\simeq&\beta_{t0} + \beta_{t2}\, \cos(2\phi_r ),
\eea
with $\Delta y_{t2} \leq \Delta y_{t0}$ for {\em positive-definite boost}. The convention $\phi_r \equiv \phi - \Psi_r$ is adopted for more compact notation where $\Psi_r$ is an event-wise reference angle that may relate to an \aa\ reaction plane {\em or not}. Monopole boost component $\Delta y_{t0}$ may be inferred from $v_2(p_t)$ data but quadrupole component $\Delta y_{t2}$ is less accessible. Monopole boost $\Delta y_{t0}$ might be associated with ``radial flow'' but may apply to only a small fraction of all  hadrons.

\section{Single-particle spectra} \label{spspec}

This appendix refers to single-particle (SP) spectra required to derive quadrupole spectra from $v_2(p_t,n_{ch})$ data. Conventional \pt-differential measure $v_2(p_t,n_{ch})$ includes SP spectrum $\bar \rho_0(p_t,n_{ch})$ in its denominator as shown in Eq.~(\ref{v2struct}). The $v_2(p_t,n_{ch})$ {\em ratio} may thus introduce a significant bias from jet contributions to the SP spectrum, aside from possible jet-related contributions to angular correlations in its numerator (``nonflow'') depending on the $v_2$ method. Unique to the NJ quadrupole is amplitude $V_2\{\rm 2D\}(p_t,n_{ch})$, which includes the quadrupole spectrum as a factor as noted in Eq.~(\ref{combfac}). To isolate quadrupole spectra from $v_{2i}(p_t,n_{ch})$ data for identified hadrons  corresponding SP spectra $\bar \rho_{0i}(p_t)$ are required.

A spectrum TCM for identified hadrons may be generated by assuming that each hadron species $i$ comprises certain fractions of soft $\bar \rho_s$ and hard $\bar \rho_h$ TCM components denoted by $z_{si}$ and $z_{hi}$  (both $\leq 1$) and assumed {\em independent of \yt} (but not of event \nch). A PID spectrum TCM for species $i$ may then be expressed as
\bea \label{pidspectcm}
\bar \rho_{0i}(y_t)
&\approx& \frac{N_{part}}{2}  \bar \rho_{sNNi} \hat S_{0i}(y_t) + N_{bin}   \bar \rho_{hNNi} \hat H_{0i}(y_t),~~~~~~
\eea
where $\bar \rho_{xNNi} = z_{xi}(n_s) \bar \rho_{xNN}$ ($x$ = $s$  or $h$) and  unit-integral model functions $\hat S_{0i}(y_t)$ and $\hat H_{0i}(y_t)$  depend on hadron species $i$. Various elements of  the PID TCM are evaluated as follows, summarized from Ref.~\cite{tompbpb}:

Geometry elements (e.g.\ $N_{part}$, $N_{bin}$) are defined in terms of hard/soft ratio $x\nu$, where $x = \bar \rho_{hNN} / \bar \rho_{sNN}$ and $\nu = 2N_{bin} / N_{part}$. $x\nu$ may in turn be inferred from ensemble-mean \mmpt\ data via the relation $\bar p_t = (\bar p_{ts} + x\nu \, \bar p_{th}) / (1 + x\nu)$, where $\bar p_{ts}$ and $\bar p_{th}$ are obtained from TCM model functions $\hat S_{0}$ and $\hat H_{0}$. That procedure assumes  \mmpt\ data have been corrected to full \pt\ spectrum acceptance.
Soft-component density $ \bar \rho_{si} =(N_{part}/2)\bar \rho_{sNNi}$ may be expressed for hadron species $i$  as
\bea \label{rhosi}
{\bar \rho_{si}} &=&  \frac{\bar \rho_{0i}}{1+\tilde z_i x\nu}
\\ \nonumber
&=&z_{si}(n_s)  \bar \rho_{s}, ~~\text{with}
\\ \nonumber
z_{si}(n_s)&\approx& \left[\frac{1 + x(n_s) \nu(n_s)}{1 + \tilde z_{i}(n_s) x(n_s) \nu(n_s)} \right]  {z_{0i}},
\eea 
where $z_{0i} = \bar \rho_{0i} /  \bar \rho_{0}$ is the fractional abundance of species $i$ relative to total density $\bar \rho_{0}$ and may correspond to statistical-model predictions, and where ratio $\tilde z_{i}(n_s) \equiv z_{hi}(n_s) / z_{si}(n_s)$ is simply proportional to hadron mass~\cite{pidpart1}. 

Figure~\ref{pbpbglaub} (left) from Ref.~\cite{tompbpb} illustrates inference of \pbpb\ geometry parameters for unidentified hadrons. Total soft density $\bar \rho_s = \bar \rho_0 / (1+ x\nu)$ is the solid curve. The lower dash-dotted curve is $\bar \rho_{sNN}$ that follows $\bar \rho_s$ up to a transition point near $\bar \rho_0 \approx 15$ and then follows a much-reduced rate of increase, taken here as $\approx 0$, following a procedure first described in Ref.~\cite{ppbpid}.

 \begin{figure}[h]
   \includegraphics[width=3.3in,height=1.65in]{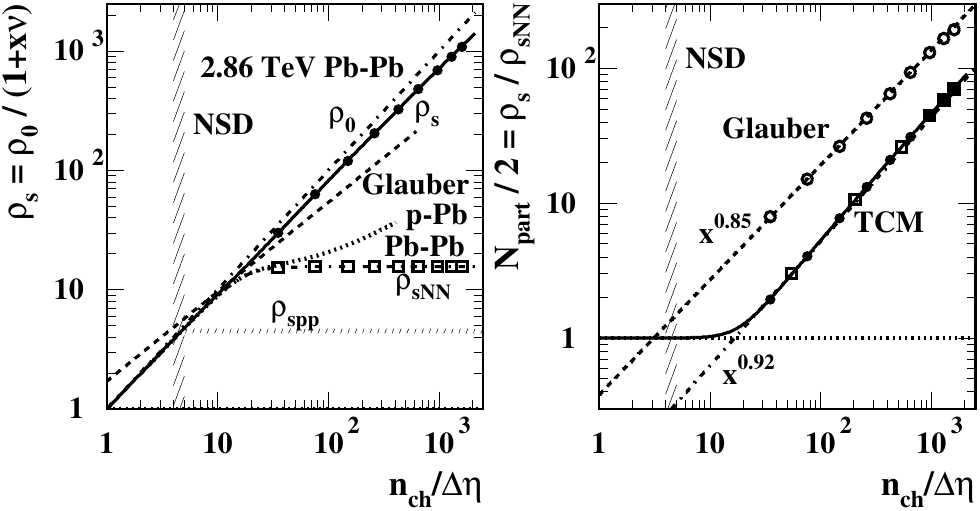}
 \caption{\label{pbpbglaub}
		Left: Soft density $\bar \rho_s = \bar \rho_0 / (1+x\nu)$ (solid dots and curve) obtained with TCM $x\nu$ derived from \mmpt\ data. The dash-dotted curve is a $\bar \rho_0$ reference. The dashed curve is obtained from a Glauber $x\nu$ trend. The lower dash-dotted and dotted curves are TCM $\bar \rho_{sNN}$ trends for \pbpb\ and \ppb\ collisions respectively.
		Right: $N_{part}/2 = \bar \rho_s / \bar \rho_{sNN}$ (TCM points and curve) derived from $\bar \rho_s$ (solid) and $x = \alpha  \bar \rho_{sNN}$ conjecture (lower dash-dotted) in (a). Open circles are Glauber values.
 }  
 \end{figure}

Figure~\ref{pbpbglaub} (right) shows $N_{part}/ 2 = \bar \rho_s /\bar \rho_{sNN} $ (solid). These TCM results may be contrasted with a classical Glauber Monte Carlo model (dashed) as reported in Ref.~\cite{pbpbcent}.  It is notable that \mmpt\ trends for \pp, \ppb\ and \pbpb\ data are {\em equivalent} up to a transition point near $\bar \rho_0 \approx 15$~\cite{tommpt}. In effect, three collision systems are identical up to that point because of {\em exclusivity}: a projectile nucleon can only interact with one target nucleon {\em at a time}, where that concept is defined in Ref.~\cite{tompbpb}. If an \mbox{A-B} collision overlap volume is too small (i.e.\ for $\bar \rho_0 < 15$) there is insufficient time for a second \nn\ collision, leading to the equivalence of the three systems. Exclusivity explains why within  that interval $N_{part}/ 2 \approx 1$ as indicated in the right panel. Similar procedures generate a matching $N_{bin}$ trend. TCM geometry parameters described above are  applied in Eq.~(\ref{v22b}) to generate the solid curve in Fig.~\ref{quadxx} (left) that passes through the \pp\ and most of the \aa\ $v_2$ data transformed to $V_2^2(n_{ch})$.

PID spectra may be rescaled by  $ \bar \rho_{si}$ in Eq.~(\ref{rhosi}) to obtain
\bea \label{xiscale}
\frac{\bar \rho_{0i}(y_t)}{ \bar \rho_{si}} 
&=& \hat S_{0i}(y_t) +  \tilde z_i(n_s) x(n_s)\nu(n_s) \hat H_{0i}(y_t,n_s)~~~
\\ \nonumber
&\equiv&   X_i(y_t)
\eea
which permits precise examination of spectrum shape variation with event \nch\ and enables isolation of PID spectrum hard components for detailed study~\cite{ppbpid,pidpart1,pidpart2}.

\begin{figure}[h]
     \includegraphics[width=3.3in]{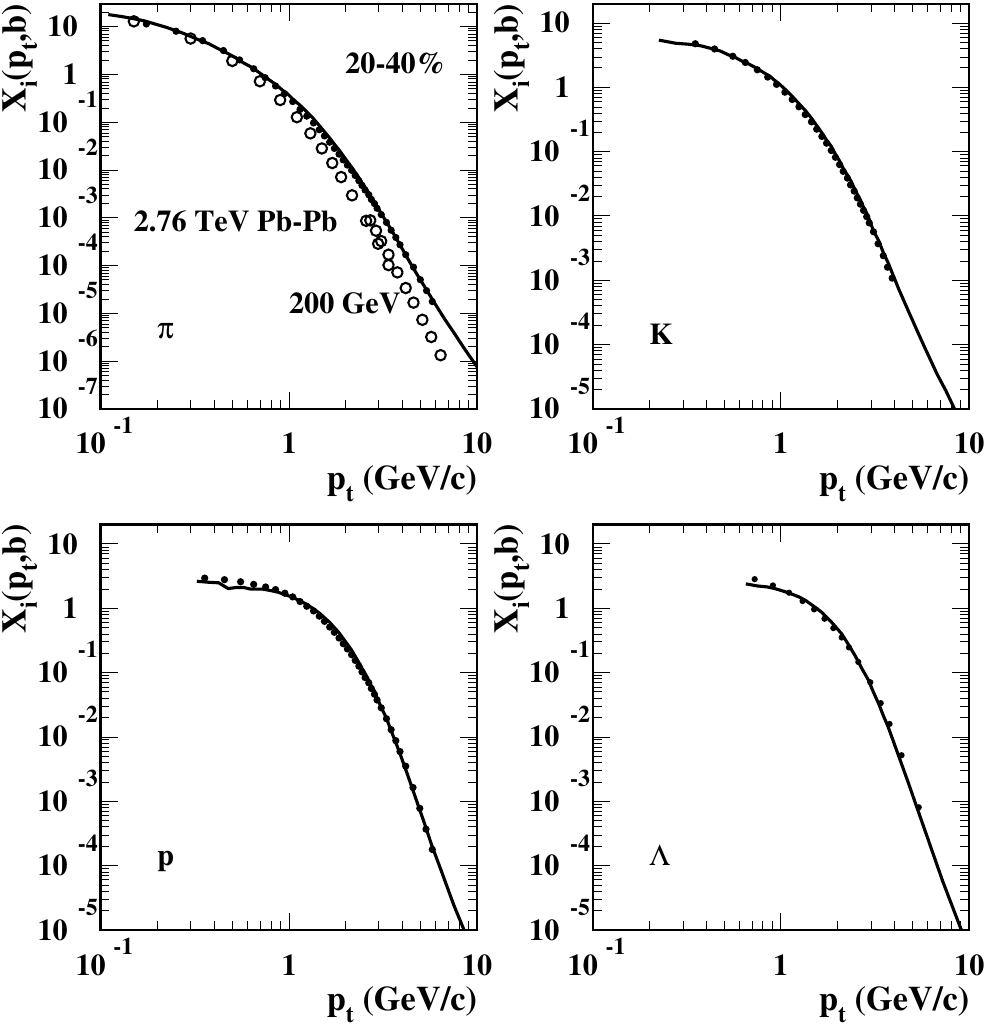}
\put(-140,213) {\bf (a)}
\put(-22,213) {\bf (b)}
\put(-140,88) {\bf (c)}
\put(-22,88) {\bf (d)}
\caption{\label{alice5}
Identified-hadron spectra for three hadron species from 20-40\%  central 2.76 TeV \pbpb\ collisions (solid curves)~\cite{alicepbpbpidspec}. Spectra $\bar \rho_{0i}(p_t)$ are rescaled to  $X_i(y_t)$ as in Eq.~(\ref{xiscale}).
The solid points are model parametrizations defined on $v_2(p_t)$ data \pt\ values from Ref.~\cite{alicev2ptb} used for the present study.  In panel (a) the open points are pion model spectra for 200 GeV \auau\ $v_2$ data used in Fig.~\ref{x2} (right). In (d) the Lambda spectrum (solid curve) is from Ref.~\cite{alicelambda}.
} 
\end{figure}    

Figure~\ref{alice5} shows SP data spectra as densities on \pt\ (solid curves) in the rescaled form $X_i(y_t)$ for identified (a) pions, (b) charged kaons, (c) protons and (d) Lambdas from 20-40\% central 2.76 TeV \pbpb\ collisions~\cite{alicepbpbpidspec}. Because jet-related contributions to $v_2$ data (nonflow) are relatively largest for peripheral and central collisions, representative $v_2(p_t,n_{ch})$ data for 20-40\% central are preferred as noted in Sec.~\ref{quadspec}. In Fig.~\ref{alice3} it is demonstrated that shape variations with \nch\ of $v_2(p_t)$ data in the form $v_2(p_t)/p_t$ are negligible compared to uncertainties, so the choice of SP spectrum centrality is not critical.

In Fig.~\ref{alice5}  dotted points approximating data (curves) are parametrized models of  \pbpb\ spectrum data defined on $v_2(p_t)$ data \pt\ values. The model values for $\rho_{0i}(y_t)$, also rescaled to $X_i(y_t)$ as in Eq.~(\ref{xiscale}), are used to determine quadrupole spectra $\rho_{2i}(y_t,\Delta y_{t0})$ in Eq.~(\ref{combfac}).

\end{appendix}



\begin{thebibliography}{99}

\bibitem{olli} J.~Y.~Ollitrault,
Phys. Rev. D \textbf{46}, 229-245 (1992).

\bibitem{perfect}   M.~Gyulassy and L.~McLerran,
  Nucl.\ Phys.\ A {\bf 750}, 30 (2005).

 \bibitem{qgp1} T.~Hirano and M.~Gyulassy,
  Nucl.\ Phys.\  A {\bf 769}, 71 (2006).

 \bibitem{qgp2} L.~P.~Csernai, J.~I.~Kapusta and L.~D.~McLerran,
  Phys.\ Rev.\ Lett.\  {\bf 97}, 152303 (2006).

\bibitem{keystone} D.~Teaney, J.~Lauret and E.~V.~Shuryak,
  Phys.\ Rev.\ Lett.\  {\bf 86}, 4783 (2001).

\bibitem{hydro2}  P.~Huovinen and P.~V.~Ruuskanen,
Ann.\ Rev.\ Nucl.\ Part.\ Sci.\  {\bf 56}, 163 (2006).

\bibitem{axialci}   J.~Adams {\it et al.}  (STAR Collaboration),
  Phys.\ Rev.\  C {\bf 73}, 064907 (2006).

\bibitem{multipoles} T.~A.~Trainor,
  J.\ Phys.\ G {\bf 40}, 055104 (2013).

\bibitem{ppquad}  T.~A.~Trainor and D.~J.~Prindle,
  Phys.\ Rev.\ D {\bf 93}, 014031 (2016).

\bibitem{anomalous}  G.\ Agakishiev, {\it et al.} (STAR Collaboration),
  Phys.\ Rev.\ C {\bf 86}, 064902 (2012).

\bibitem{davidhq}  D.~T.~Kettler  (STAR collaboration),
  Eur.\ Phys.\ J.\  C {\bf 62}, 175 (2009).

\bibitem{aliceptfluct}  T.~A.~Trainor,
  Phys.\ Rev.\ C {\bf 92}, 024915 (2015).

\bibitem{alicetomspec}T.~A.~Trainor,
J. Phys. G \textbf{44}, no.7, 075008 (2017)

\bibitem{hardspec}  T.~A.~Trainor,
  Int.\ J.\ Mod.\ Phys.\  E {\bf 17}, 1499 (2008).

\bibitem{quadspec}   T.~A.~Trainor,
  Phys.\ Rev.\  C {\bf 78}, 064908 (2008).

\bibitem{statmodel}  F.~Becattini, M.~Gazdzicki and J.~Sollfrank,
  Eur.\ Phys.\ J.\  C {\bf 5}, 143 (1998).

\bibitem{poskvol} A.~M.~Poskanzer and S.~A.~Voloshin,
Phys. Rev. C \textbf{58}, 1671-1678 (1998).

\bibitem{bevalac} H. A. Gustafsson, {\em et al.}, Phys.\ Rev.\ Lett.\ {\bf 52}, 1590 (1984). 

\bibitem{njquad}  T.~A.~Trainor,
arXiv:1610.06256.

\bibitem{cumulant} A.~Bilandzic, R.~Snellings and S.~Voloshin,
Phys. Rev. C \textbf{83}, 044913 (2011).

\bibitem{subevents} M.~Aaboud \textit{et al.} (ATLAS),
Phys. Rev. C \textbf{97}, no.2, 024904 (2018).

\bibitem{ppbpid}  T.~A.~Trainor,
J.\ Phys.\ G \textbf{47}, no.4, 045104 (2020).

\bibitem{ppprd} J.~Adams {\it et al.}  (STAR Collaboration),
  Phys.\ Rev.\  D {\bf 74}, 032006 (2006).

\bibitem{tompbpb} T.~A.~Trainor,
arXiv:2510.05314.

\bibitem{noblast} T.~A.~Trainor,
arXiv:2206.07791.

\bibitem{cooperfrye}  F.~Cooper and G.~Frye,
  Phys.\ Rev.\  D {\bf 10}, 186 (1974).

\bibitem{v2pions}  C.~Adler {\it et al.}  (STAR Collaboration),
  Phys.\ Rev.\  C {\bf 66}, 034904 (2002).

\bibitem{v2strange}
J.~Adams {\it et al.}  (STAR Collaboration),
  Phys.\ Rev.\ Lett.\  {\bf 92}, 052302 (2004).

\bibitem{newstarpion} B.~I.~Abelev {\it et al.} (STAR Collaboration),
Phys.\ Rev.\ C {\bf 77}, 054901 (2008).

\bibitem{luzrat} M.~Luzum and P.~Romatschke,
Phys. Rev. C \textbf{78}, 034915 (2008).

\bibitem{rom} P.~Romatschke and U.~Romatschke,
  Phys.\ Rev.\ Lett.\  {\bf 99}, 172301 (2007).

\bibitem{shen} C.~Shen, U.~Heinz, P.~Huovinen and H.~Song,
Phys. Rev. C \textbf{84}, 044903 (2011).

\bibitem{alicev2ptb} B.~B.~Abelev \textit{et al.} [ALICE],
JHEP \textbf{06}, 190 (2015).

\bibitem{2004} J.~Adams {\it et al.}  (STAR Collaboration),
  Phys.\ Rev.\  C {\bf 72}, 014904 (2005).

\bibitem{tomglauber}  T.~A.~Trainor,
arXiv:1801.05862
 
\bibitem{alicev2b}  K.~Aamodt {\it et al.} (ALICE Collaboration),
Phys.\ Rev.\ Lett.\  {\bf 105}, 252302 (2010).

\bibitem{v2ptb} D.~T.~Kettler, D.~J.~Prindle and T.~A.~Trainor,
  Phys.\ Rev.\ C {\bf 91}, 064910 (2015).

\bibitem{noelliptic} T.~A.~Trainor, D.~T.~Kettler, D.~J.~Prindle and R.~L.~Ray,
 J.\ Phys.\ G  {\bf 42}, 025102 (2015).

\bibitem{alicepbpbpidspec}  J.~Adam {\it et al.} (ALICE Collaboration),
Phys.\ Rev.\ C {\bf 93}, no. 3, 034913 (2016).

\bibitem{rudy}  R.~C.~Hwa {\em et al.,} 
Phys.\ Rev.\ C {\bf 70}, 024905 (2004).

\bibitem{duke} R.~J.~Fries {\em et al.,} 
Phys.\ Rev.\ C {\bf 68}, 044902 (2003).

\bibitem{tamu} V.~Greco {\em et al.,} 
Phys.\ Rev.\ Lett.\ {\bf 90}, 202302 (2003).

\bibitem{cmsridge}
V.~Khachatryan \textit{et al.} (CMS),
JHEP \textbf{09}, 091 (2010).

\bibitem{kolk} N.~van der Kolk,
CERN-THESIS-2012-019.

\bibitem{cmsabv2} A.~M.~Sirunyan \textit{et al.} (CMS),
Phys. Rev. Lett. \textbf{120}, no.9, 092301 (2018).

\bibitem{tomalicempt}  T.~A.~Trainor,
  Phys.\ Rev.\ C {\bf 90}, no. 2, 024909 (2014)

\bibitem{alicerho0} K.~Aamodt \textit{et al.} (ALICE),
Phys. Rev. Lett. \textbf{106}, 032301 (2011).
 
\bibitem{tomexclude}  T.~A.~Trainor,
arXiv:1801.06579.

\bibitem{powerlaw} T.~A.~Trainor and D.~J.~Prindle,
arXiv:hep-ph/0411217.

\bibitem{jetspec2}  T.~A.~Trainor,
Phys.\ Rev.\ D  {\bf 89}, 094011 (2014).

\bibitem{squeezeout} H.~H.~Gutbrod, K.~H.~Kampert, B.~Kolb, A.~M.~Poskanzer, H.~G.~Ritter, R.~Schicker and H.~R.~Schmidt,
Phys. Rev. C \textbf{42}, 640-651 (1990).

\bibitem{pbm} P.~Braun-Munzinger, K.~Redlich and J.~Stachel,
nucl-th/0304013 [nucl-th].

\bibitem{uliflow} U.~Heinz and R.~Snellings,
Ann. Rev. Nucl. Part. Sci. \textbf{63}, 123-151 (2013).

\bibitem{gribov}  Y.~L.~Dokshitzer and D.~E.~Kharzeev,
Ann.\ Rev.\ Nucl.\ Part.\ Sci.\  {\bf 54}, 487 (2004).

\bibitem{pidpart2} T.~A.~Trainor,
arXiv:2112.12330.

\bibitem{fragevo}    T.~A.~Trainor,
  Phys.\ Rev.\  C {\bf 80}, 044901 (2009).

\bibitem{songperliq} H.~Song, S.~A.~Bass, U.~Heinz, T.~Hirano and C.~Shen,
Phys. Rev. Lett. \textbf{106}, 192301 (2011)
[erratum: Phys. Rev. Lett. \textbf{109}, 139904 (2012)].

\bibitem{ppbnmf} T.~A.~Trainor,
arXiv:2304.02170.

\bibitem{pidpart1} T.~A.~Trainor,
arXiv:2112.09790.

\bibitem{pppid} T.~A.~Trainor,
arXiv:2210.05877.

\bibitem{alicespec} T.~A.~Trainor,
J. Phys. G \textbf{44}, no.7, 075008 (2017).

\bibitem{tcmcompare} T.~A.~Trainor,
arXiv:2401.03290.

\bibitem{mbjets} T.~A.~Trainor,
arXiv:1701.07866.

\bibitem{lisa} F.~Retiere and M.~A.~Lisa,
Phys. Rev. C \textbf{70}, 044907 (2004).

\bibitem{pbpbcent}  B.~Abelev {\it et al.} (ALICE Collaboration),
Phys.\ Rev.\ C {\bf 88}, no. 4, 044909 (2013).

\bibitem{tommpt} T.~A.~Trainor,
arXiv:1708.09412.

\bibitem{alicelambda} B.~B.~Abelev \textit{et al.} (ALICE),
Phys. Rev. Lett. \textbf{111}, 222301 (2013).



\end{thebibliography}
\end{document}